\documentclass[showpacs,twocolumn,aps,prb,floatfix,superscriptaddress,noshowpacs]{revtex4}
\usepackage[mathlines]{lineno}% Enable numbering of text and display math

\usepackage{amsmath,amssymb,graphicx,bm,color,mathrsfs,verbatim,epstopdf,dcolumn,dsfont}
\usepackage{natbib}
\usepackage{ulem}

\usepackage{hyperref}
\hypersetup{
	colorlinks   = true
}

\newcommand{\be}{\begin{equation}}
\newcommand{\ee}{\end{equation}}

\begin{document}

%\preprint{APS/123-QED}

\title{Heating and many-body resonances in a periodically driven two-band system}%

%\author{A permutation of}

\author{Marin Bukov}
\email{mbukov@bu.edu}
\affiliation{Department of Physics, Boston University, 590 Commonwealth Ave., Boston, MA 02215, USA}

\author{Markus Heyl}
\affiliation{Physik Department, Technische Universit\"at M\"unchen, 85747 Garching, Germany}

\author{David A. Huse}
\affiliation{Physics Department, Princeton University, Princeton, NJ 08544, USA, and Institute for Advanced Study, Princeton, NJ 08540, USA}

\author{Anatoli Polkovnikov}
\affiliation{Department of Physics, Boston University, 590 Commonwealth Ave., Boston, MA 02215, USA}

\date{\today}

\begin{abstract}
We study the dynamics and stability in a strongly interacting resonantly driven two-band model. Using exact numerical simulations, we find a stable regime at large driving frequencies where the time evolution is governed by a local Floquet Hamiltonian that is approximately conserved out to very long times. For slow driving, on the other hand, the system becomes unstable and heats up to infinite temperature.  While thermalization is relatively fast in these two regimes (but to different ``temperatures''), in the crossover between them we find slow nonthermalizing time evolution: temporal fluctuations become strong and temporal correlations long lived.  Microscopically, we trace back the origin of this nonthermalizing time evolution to the properties of rare Floquet many-body resonances, whose proliferation at lower driving frequency removes the approximate energy conservation, and thus produces thermalization to infinite temperature.
\end{abstract}


%\pacs{67.85.Pq, 71.10.Fd}

\maketitle

\section{\label{sec:intro}Introduction}

Periodically-driven cold atomic systems have recently proven indispensable for engineering models otherwise inaccessible in static systems. This includes gauge fields~\cite{jaksch_03,mueller_04,eckardt_10,creffield_11,kolovsky_11,struck_11,struck_12,hauke_12,struck_13,aidelsburger_13,miyake_13,atala_14,aidelsburger_14,kennedy_15,flaeschner_15}, topological~\cite{oka_09,kitagawa_11,jotzu_14,aidelsburger_14,grushin_14,anisimovas_15,verdeny_13,verdeny_15} and spin-dependent~\cite{jotzu_15} bands as well as dynamical localisation and stabilisation~\cite{dunlap_86,dunlap_88,lignier_07,sias_07,eckardt_09,zenesini_09,creffield_10}, correlated tunnelling~\cite{meinert_16}, and Floquet topological pumps~\cite{rudner_13,lindner_16}, just to name a few. The state of the art in the field currently identifies heating as a major experimental challenge~\cite{aidelsburger_13,miyake_13,aidelsburger_14,kennedy_15,jotzu_14,weinberg_15}, yet the respective mechanisms are not fully understood. In particular, due to the large driving frequencies involved in the process of Floquet engineering, transitions to higher bands become energetically possible, raising the immediate question as to whether this inevitably leads to indefinite heating. If so, however, are there windows in parameter space in which heating is suppressed or can be controlled?

\begin{figure}[h!]
	\includegraphics[width=0.6\columnwidth]{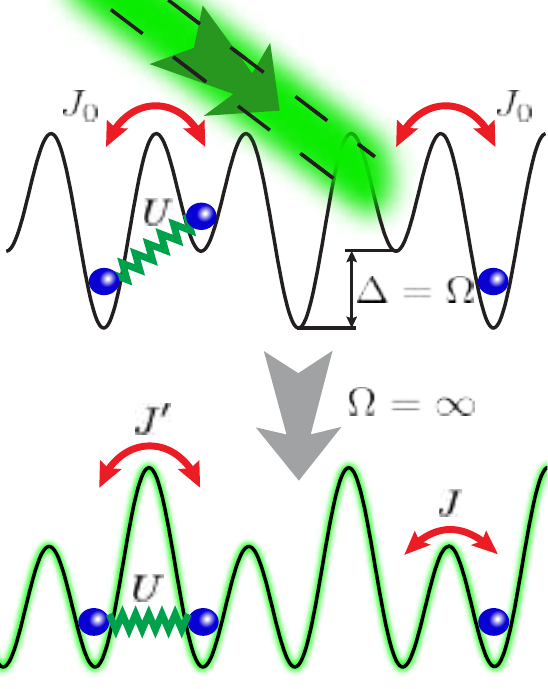}
	\caption{\label{fig:model} Floquet realisation of the interacting SSH model: the non-driven system represents a two-band model, coupled resonantly by a strong periodic drive. As a result, the ground state of the infinite-frequency Floquet Hamiltonian features an interesting topological phase.}
\end{figure}

In this study we demonstrate that there are regimes where a resonant coupling of two bands
does not produce strong heating on the experimentally accessible time scales even in the presence of strong interactions. In particular, for sufficiently large driving frequencies, we find evidence, on the basis of exact numerical simulations, that heating is perturbatively weak and, therefore, controllable. On the other hand, decreasing the driving frequency below a crossover scale $\Omega^*$ of order the single-particle bandwidth of the noninteracting system, our model exhibits strong heating.  In the crossover regime, we find a range of driving frequencies where the system displays slow non-thermalising time-evolution: we observe long-lived temporal fluctuations and correlations which do not decay on the experimentally relevant time scales.  We argue that this non-thermalising behavior arises due to rare Floquet many-body resonances.

Next to the study of thermalisation in periodically-driven systems from a theory point of view, the second purpose of our work is to study the onset of heating at the experimentally-observable times. Present-day experiments with cold atoms performed in the high-frequency regime report heating after a few hundred driving cycles~\cite{aidelsburger_14}. Interestingly, heating seems to be more pronounced in bosonic rather than fermionic~\cite{jotzu_14,jotzu_15} systems, presumably due to the unbounded character of the on-site Hilbert space dimension. A Density Matrix Renormalisation Group (DMRG) study in the weakly-interacting, periodically-driven Bose-Hubbard chain~\cite{poletti_11} found that heating is indeed suppressed at the large frequencies necessary to create novel Floquet Hamiltonians~\cite{sias_07,lignier_07,eckardt_05,zenesini_09}. Moreover, the existence of long-lived prethermal Floquet steady states has been predicted~\cite{kuwahara_15,abanin_15_2} and confirmed numerically~\cite{canovi_15,bukov_15_prl}. In this paper we study a minimal model of two resonantly-coupled bands, and find that heating is suppressed at large frequencies allowing for controlled Floquet engineering.

%%%%%%%%%%%%%%%%%%%%%%%%%%%%%

\section{\label{sec:model}Model}

Consider a system of interacting hardcore bosons satisfying three main properties, as illustrated in Fig.~\ref{fig:model}:
(i) the non-driven system represents a two-band model, (ii) the periodic drive couples resonantly the two bands,
and (iii) the ground state of the infinite-frequency Floquet Hamiltonian exhibits an interesting topological phase.
The model can be equivalently mapped with a Jordan-Wigner transformation to spinless fermions, but we choose to present it here as hardcore bosons.

Concretely, the full dynamics is encoded in the Hamiltonian $H(t) = H_0 + H_\mathrm{drive}(t)$ with $H_0$ the non-driven two-band model:
\begin{eqnarray}
H_0 &=& -J_0\sum_{j=1}^{L-1}\left( a^\dagger_{j+1}a_j+ \text{h.c.}\right) -\frac{\Delta}{2}\sum_{j=1}^L (-1)^jn_j  \nonumber\\
&& + U\sum_{j=1}^{L-1} \left(n_j-\frac{1}{2}\right)\left( n_{j+1} - \frac{1}{2}\right).
\label{eq:SSH_nondriven}
\end{eqnarray}
Here the operator $a^\dagger_j$ creates a hardcore boson at $j=1,\dots,L$ with $L$ the total number of lattice sites,
$n_j=a_j^\dag a_j$ the number operator, $J_0$ denotes the bare hopping amplitude, $\Delta$ -- the strength of a staggered potential, and $U$ -- the interaction strength. We limit the discussion to half filling with $L$ even. When $J_0$, $\Delta$ and $U$ are all nonzero this model is non-integrable~\cite{benenti_09}. The non-interacting model has two bands, separated by the gap $\Delta$. The periodic drive is
\begin{eqnarray}
H_\text{drive}(t) &=& f(t)\;\sum_{j=1}^L \left[ \frac{A}{2}(-1)^j - \delta A\; j\right]n_j,
\label{eq:SSH_drive}
\end{eqnarray}
with the time-periodic step function $f(t) = \mathrm{sign}[\cos(\Omega t)]$, $A$ -- the amplitude of the modulated superlattice, $\delta A$ -- the amplitude of the shaken external field, and $T = 2\pi/\Omega$ -- the driving period. Compared to a monochromatic driving, $f(t)$ contains higher harmonics of $\Omega$ which, however, we checked does not change the  phenomena discussed below. Therefore, in the followig, using the relation $\Omega=2\pi/T$, we shall refer to $\Omega$ as the frequency of the drive.

To study the amount of heating [i.e.~excess energy produced in the system] in the tight-binding limit, it is enough to consider stroboscopic dynamics. Mathematically, this follows from Floquet's theorem, according to which the evolution operator is given by $U(t,0) = P(t)\exp\left(-iH_F t\right)$. Since the unitary operator which governs the fast motion~\cite{bukov_14} is periodic, $P(t+T) = P(t)$, it suffices to look at the system at stroboscopic times when $U(lT,0) = \exp\left(-iH_F lT\right)$. Intuitively, one needs to close a full driving cycle before comparing the value of the energy to the initial one. Only then can one make a statement about the amount of energy pumped into the system by the drive. However, if one of the parameters in the model, e.g.~the driving amplitude, is being changed in the presence of the drive~\cite{pweinberg_15}, or if the system is not completely described by a tight-binding model~\cite{bilitewski_15}, then one needs to take into account the heating effects due to the change of the $P$-operator as well. 

In the following, we always set $\Delta = \Omega$, which resonantly couples and mixes the two bands of the non-driven Hamiltonian $H_0$~\cite{goldman_14_res,bukov_15_SW}. In the high-frequency regime, the effective Floquet Hamiltonian $H_F$ governing the stroboscopic time-evolution of the system,
\begin{equation}
U_F = \mathcal{T}_t\mathrm{exp}\left(-i\int_0^TH(t)\mathrm{d}t\right) = e^{-iH_F T},
\label{eq:U_F}
\end{equation}
can be found with the help of an inverse-frequency expansion~\cite{rahav_03_pra,rahav_03,goldman_14,bukov_14,goldman_14_res,eckardt_15,mikami_15}. We refer to $U_F$ as the Floquet operator. Since we choose the driving amplitude $A$ as well as the superlattice potential $\Delta$ to be on the order of the driving frequency $\Omega$, the time-average has to be performed in the rotating frame~\cite{bukov_14}. In the infinite-frequency limit the Floquet Hamiltonian reads:
\begin{equation}
\! H_F^{(0)} \! = \! \sum_{j=1}^{L-1} \!-\! J_j \!\left( \!a^\dagger_{j+1} a_{j}\! + \! \text{h.c.} \!\right)\! + \! U\! \!\left( \! n_{j+1}\!-\!\frac{1}{2}\! \right)\! \!\left( \!n_{j}\!-\!\frac{1}{2} \!\right) \!,
\label{eq:HF0}
\end{equation}
where the drive-renormalised hopping elements are $J_j = J = J_0 \chi(\zeta-\delta \zeta)$ for $j$ odd, and $J_j = J' = J_0 \chi(\zeta+\delta\zeta)$ for $j$ even. Here $\chi(x) =  2x\pi^{-1}\cos(\pi x/2)/(1-x^2)$, $\zeta = A/\Omega$, and $\delta\zeta = \delta A/\Omega$. Thus, $H_F^{(0)}$ realizes the Su-Schrieffer-Heeger (SSH) model including additionally nearest-neighbour interactions. When $J\neq J'$ and $U\neq 0$, this model is quantum chaotic with GOE level statistics, see~App.~\ref{app:level_statistics}.  For $U=0$, the system features two topological bands whenever $J\neq J'$, separated by a gap of energy width $2|J-J'|$. Notice how the topological gap is opened solely due to the drive, in close analogy with the experimental realizations of the Harper-Hofstadter model and the Haldane model in two-dimensions~\cite{aidelsburger_13,miyake_13,aidelsburger_14,jotzu_14,kennedy_15}.

In analogy to recent experiments we study the following general protocol. We initialize the system in the ground state $|\psi\rangle$ of the topological infinite-frequency Floquet Hamiltonian $H_F^{(0)}$ which, to a good accuracy, can be also generated experimentally via adiabatic state preparation~\cite{aidelsburger_13,miyake_13,aidelsburger_14,jotzu_14,kennedy_15,pweinberg_15}. Heating effects due to the adiabatic state preparation in the presence of the drive are discussed elsewhere, cf.~Ref.~\onlinecite{pweinberg_15}, where it is demonstrated that at high driving frequencies one can generally prepare ground states of Floquet Hamiltonians with a high, though not perfect, fidelity. The subsequent dynamics, which we are interested in, is generated by the full time-dependent Hamiltonian, see Eqs.~\ref{eq:SSH_nondriven},~\ref{eq:SSH_drive} and~\ref{eq:U_F}. To study the dynamics numerically, (i) we calculate the exact evolution w.r.t.~the Hamiltonian $H(t)$ using a Lanczos algorithm with full reorthogonalisation based on Krylov's method, which allows us to study the first several thousand driving periods for system sizes up to $L=20$. Since we are interested in stroboscopic evolution, (ii) we also compute the exact Floquet operator $U_F$ and apply exact diagonalisation (ED): projecting the initial state onto the Floquet basis allows us to directly reach the infinite-time limit for system sizes up to $L=16$, by means of a quench to the diagonal ensemble of $U_F$. A detailed comparison between the two methods, as well as the system-size dependence of the results discussed below is presented in App.~\ref{app:finite_size_scaling}.

%%%%%%%%%%%%%%%%%%%%%%%%%%%%%
%	Heating

\section{\label{sec:heating}Heating}

After having specified the details of the model system and the protocol of the drive, it is the purpose of the following section to study the heating dynamics as a function of the driving frequency $\Omega$. Specifically, we will characterize the heating on the basis of the energy absorbed by the system from the drive in Sec.~\ref{subsec:E_absorption}, as well as the half-chain entanglement entropy in Sec.~\ref{subsec:ent_entropy}. Last but not least, in Sec.~\ref{subsec:U_dep}, we briefly discuss the dependence of heating on the interaction strength.

\subsection{\label{subsec:E_absorption}Energy Absorption}

Let us begin the study of the heating dynamics by looking at the energy of the system. In analogy to experiments, where it is the Floquet-engineered infinite-frequency Hamiltonian $H_F^{(0)}$ that is the prime object of interest, we characterize  heating by measuring the energy via $H_F^{(0)}$ in the time-evolved state. Specifically, we calculate the stroboscopic evolution~\cite{bukov_14_pra} of the energy density $\mathcal{E}_\psi$ of $H_F^{(0)}$:
\be
\mathcal{E}_\psi(lT)= \frac{1}{L}\langle \psi| H_F^{(0)} (lT)|\psi \rangle,
\ee
with $l\in\mathbb{N}$, and the time-dependence of $H_F^{(0)}(T) = U^\dagger_FH_F^{(0)}U_F$ is understood in the Heisenberg picture. While in the infinite-frequency limit $\mathcal{E}_\psi(lT)=\mathcal{E}_\psi(0)=\mathrm{const.}$ and heating is absent, at finite $\Omega$ the system will be driven out of the initial ground-state manifold and will increase its energy.

\begin{figure}
	\includegraphics[width=1.0\columnwidth]{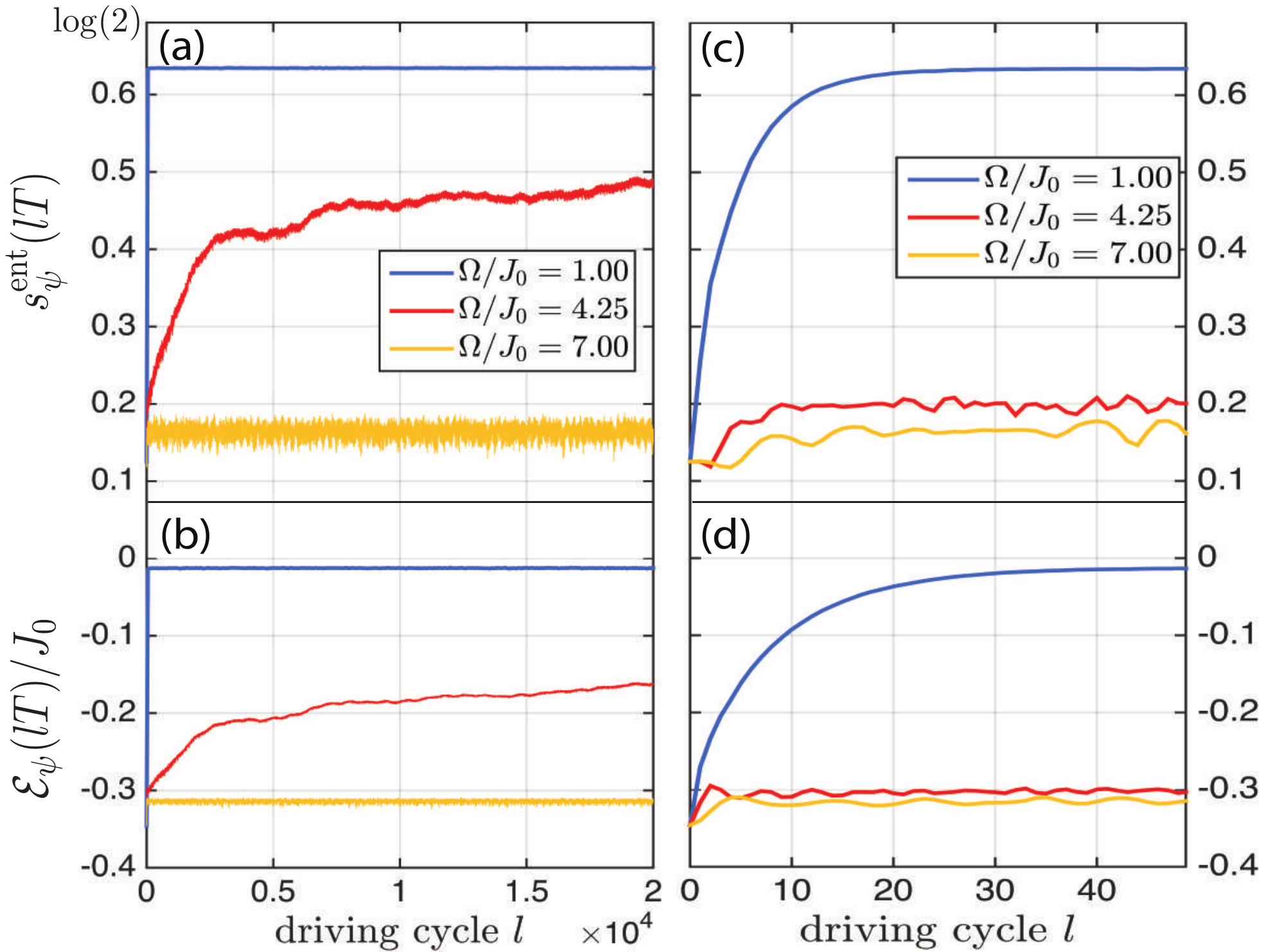}
	\caption{\label{fig:energy_entropy} Short and long-time stroboscopic dynamics of the entropy density (a) and (c), and the energy density (b) and (d). The linewidths in (a) and (b) show the size of temporal fluctuations.
		% up to scale.
		The parameters are $U/J_0=1$, $\zeta = 0.6$, $\delta\zeta = 0.12$, and $L=20$, which in the high-frequency limit gives $J'/J_0 = 0.41$ and $J/J_0=0.29$. }
\end{figure}

Depending on the magnitude of the driving frequency, we identify two different regimes, separated by a crossover scale $\Omega^*$, see~Figs.~\ref{fig:energy_entropy},~\ref{fig:U_dep}. A quantitative analysis of $\Omega^\ast$ for small interactions can be found in Sec.~\ref{subsec:U_dep} and specifically in Eq.~\ref{eq:omegaStar}. For $\Omega \ll \Omega^*$ the system heats up quickly close to an infinite-temperature state where all states of $H_F^{(0)}$ are occupied with equal probability. For $\Omega \gg \Omega^*$, on the other hand, heating is weak and the evolution is well-approximated by the local Floquet Hamiltonian $H_F^{(0)}$. It is interesting, from the point of view of both theory and experiment, to study the full crossover from the stable to the unstable regime as a function of the driving frequency. For that purpose, we introduce a normalised heating $\overline{Q}_\psi$, which measures the amount of energy absorbed by the system from the drive:
\begin{equation}
\overline{Q}_\psi = \frac{\overline{\mathcal{E}}_\psi-\mathcal{E}_\psi(0)}{\mathcal{E}_{\beta=0}-\mathcal{E}_\psi(0)}.
\end{equation}
$\overline{Q}_\psi$ interpolates continuously between absence of heating, where~$\overline{Q}_\psi=0$, and heating to infinite temperature, where $\overline{Q}_\psi=1$, see also Fig.~\ref{fig:U_dep}. Here, $\overline{\mathcal{E}}_\psi = \lim_{N_T \to \infty} N_T^{-1} \sum_{l=1}^{N_T} \mathcal{E}_\psi(lT) $ is the stroboscopic time average of $\mathcal E_\psi(lT)$, while $\mathcal{E}_{\beta=0}$ is the infinite-temperature average, which is close to the centre of the many-body band, up to $L^{-1}$-corrections [$\mathcal{E}_{\beta=0}=-U/(4L)$ for half-filling].  We calculated the long-time limit from a time average of the stroboscopic evolution over the last $4\times 10^3$ of $5\times 10^3$ total driving periods obtained via the aforementioned Lanczos algorithm. We checked that nonzero initial temperatures do not change the physical picture, see App.~\ref{app:finite-Temperature}. A more detailed analysis of finite-size effects is given in App.~\ref{app:finite_size_scaling}. While we find that the results appear to be only weakly sensitive to increasing $L$, finite-size effects become most pronounced in the vicinity of the crossover scale $\Omega^*$.

To understand the origin of this behaviour, we show both the short-term evolution, Fig.~\ref{fig:energy_entropy}d, relevant for present-day experiments with cold atoms, as well as the longer-term behaviour, Fig.~\ref{fig:energy_entropy}b, which allows us to make a statement about energy absorption in the longer run. For $\Omega \gg \Omega^*$, the energy density stays at a value near the ground state of $H_F^{(0)}$, which is perturbatively controlled by the inverse frequency $\Omega^{-1}$ and becomes vanishingly small upon including higher-order $\Omega^{-1}$--corrections to the approximate Floquet Hamiltonian cf.~App.~\ref{app:Floquet corrections}. Therefore, in this regime the dynamics is completely stable on the experimentally relevant time scales for the numerically simulated system sizes. It follows that heating can be well-controlled making this parameter regime particularly suitable for Floquet engineering.

Conversely, for $\Omega \ll \Omega^*$, the energy absorption becomes strong which leads to fast heating with the energy quickly approaching its infinite-temperature value. Hence, the system is unstable and experiments in this regime are rendered uncontrollable. It is, thus, crucial to acquire a better understanding of the frequency-dependence of the onset of heating. Interestingly, in the vicinity of the crossover scale $\Omega\approx \Omega^*$, the dynamics changes its character completely. Although the system still heats up, [but not to infinite temperature for finite system size $L$], the time scales become so extended that the final relaxation cannot be resolved within the studied $2\times 10^4$ driving cycles, see Fig.~\ref{fig:energy_entropy}. The origin of this substantially slowed down dynamics we analyze in more detail in Sec.~\ref{sec:ergodicity} and Sec.~\ref{sec:MB_res} where we also give explanations for the microscopic mechanism behind this unexpected behaviour.

\subsection{\label{subsec:ent_entropy} Entanglement Entropy}

The two heating regimes separated by the crossover scale $\Omega^\ast$ are also clearly identifiable from the analysis of the entanglement entropy density of half the chain:
\[
s^\mathrm{ent}_\psi(lT) = -\frac{1}{L/2}\mathrm{Tr}_{B} \left[ \rho_B(lT) \log \rho_B(lT) \right],
\]
where $B$ denotes the set of the first $L/2$ lattice sites and $\rho_B(lT)$ -- the reduced density matrix of $B$ after $l$ periods. The behavior of $s^\mathrm{ent}_\psi(lT)$ as a function of time is plotted in Fig.~\ref{fig:energy_entropy}a,c, and clearly shows the same three qualitatively different behaviours already revealed by $\mathcal{E}_\psi$. (i) At high frequencies [compared to the bare model parameters], the production of entanglement entropy remains low. The non-zero tail most likely has a two-fold origin: part of it comes from the non-zero entanglement entropy of the Floquet ground state [cf.~value at $l=0$], while the dynamically produced entanglement is due to the small temperature resulting from the energy density injected in the system by abruptly turning on the periodic drive. (ii), in the crossover, $\Omega\approx\Omega^*$, the dynamics is again found to be slow. Notice that extremely long observation times are required to fully resolve the crossover regime. (iii), for $\Omega\ll\Omega^*$ the entanglement entropy grows quickly to its infinite-temperature value of $\log(2)$ per lattice site, signalling that an infinite-temperature state is reached. From a fundamental point of view, however, $s^\mathrm{ent}_\psi$ is an even stronger indicator of the described phenomenology, since it contains information about the entire reduced density matrix. We note that the generation of entanglement entropy in integrable periodically-driven systems was studied in Refs.~\onlinecite{sen_15,russomanno_16}, while its thermalisation in a non-integrable spin chain was discussed in Ref.~\onlinecite{zhang_15}.

\subsection{\label{subsec:U_dep} Heating Dependence on the Interaction Strength}

\begin{figure}[h!]
	\includegraphics[width=0.8\columnwidth]{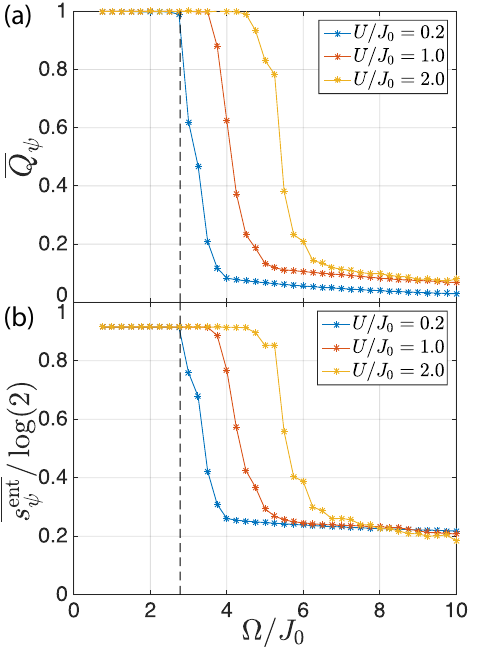}
	\caption{\label{fig:U_dep} Dependence of the crossover regime on the interaction strength: excess heat (a) and excess entanglement entropy density (b). Unity on the vertical axis corresponds to an infinite-temperature state, while zero -- to no heating. The parameters are $\zeta = 0.6$, $\delta\zeta = 0.12$, and $L=20$, which in the high-frequency limit gives $J'/J_0 = 0.41$ and $J/J_0=0.29$. }
\end{figure}

It is interesting to briefly mention the heating dependence on the interaction strength $U$. Intuitively, one would expect that a strongly interacting nonintegrable system subject to a non-energy-conserving driving protocol can easily redistribute the absorbed energy among many states due to the presence of enhanced collisions. Contrary to this naive expectation, for the system sizes up to $L=20$, we find that this does not happen for large driving frequencies, cf.~Fig.~\ref{fig:U_dep}. Instead, we find that the crossover scale $\Omega^* = \Omega^*(J_0,U,A)$ slowly shifts to higher frequencies with increasing the interaction strength $U$. Notice that in the high-frequency regime $\Omega\gg\Omega^*$, for $U/J_0=2$ the system is already strongly interacting due to the dynamically suppressed effective hopping matrix elements of the relevant infinite-frequency Floquet Hamiltonian: $U/J,U/J'\approx 10$.

For small $U$, the energy absorption appears once the full bandwidth of the single-particle Floquet Hamiltonian exceeds $\Omega/2$.  This enables heating via the basic two-particle-two-hole interaction process where two particles from the very bottom of the lower
single-particle band get scattered to the very top of the upper band. As a consequence, asymptotically for weak interactions, heating starts to occur whenever such a single-particle resonance is available. This implies the following dependence of the crossover scale for weak interactions up to corrections vanishing asymptotically for $U/J_0 \to 0$:
\begin{equation}
  \Omega^\ast = 4 \big( J + J' \big) + \mathcal{O}(U).
  \label{eq:omegaStar}
\end{equation}
Beyond the weakly interacting limit, we observe that the onset of heating $\Omega^*$ is shifted to larger values for increasing $U$, see Fig.~\ref{fig:U_dep}, presumably because higher-order processes become the effective sources of heating.  Here we do not consider the limiting case of $J_0\ll \Omega\sim U$, which can be treated using the generalised Schrieffer-Wolff transformation for periodically-driven systems~\cite{bukov_15_SW}.

%%%%%%%%%%%%%%%%%%%%%%%%%%%%%
%	Ergodicity

\section{\label{sec:ergodicity}Thermalisation -- temporal fluctuations and correlations}

\begin{figure}
	\includegraphics[width=\columnwidth]{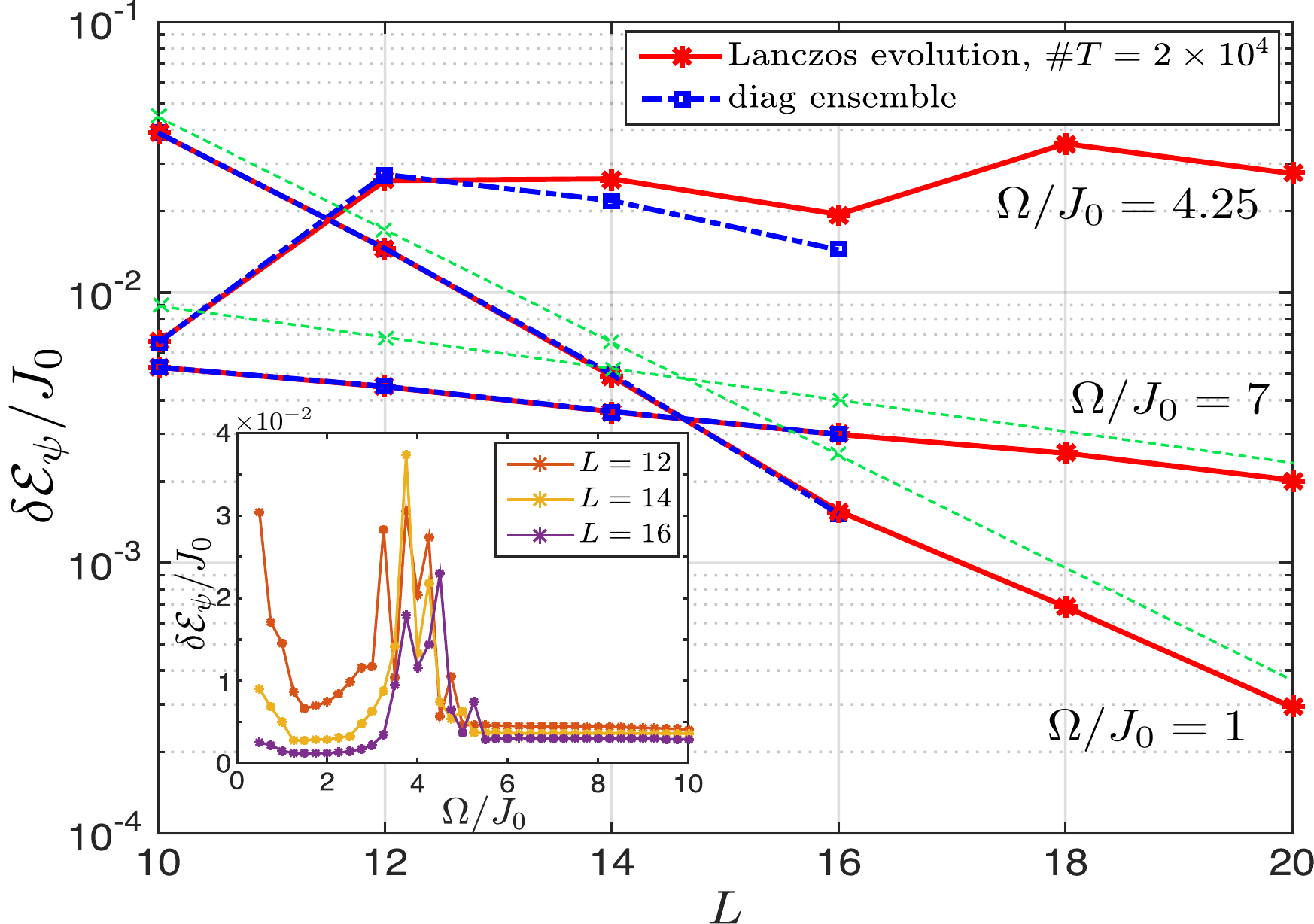}
	\caption{\label{fig:fluctuations} Energy density fluctuations as a function of the system size.  The dashed green lines show the numerical data for $\exp(-S_{\psi,d}^F/2)$ at $\Omega/J_0=1$ and $\Omega/J_0=7$ up to $L=16$, and are extrapolated for $L>16$. Here $S_{\psi,d}^F$ is the Floquet diagonal entropy, cf.~App.~\ref{app:definitions_obs}.  Inset: frequency dependence of the fluctuations at infinite time obtained using ED.  The parameters are $U/J_0=1$, $\zeta = 0.6$, $\delta\zeta = 0.12$. }
\end{figure}

In the previous Section we studied the heating dynamics as a function of driving frequency. As a main observation, we identified an extended crossover region with extremely slow dynamics that separates the regimes of unstable heating from the stable region where the dynamics is approximately governed by the desired infinite-frequency Floquet Hamiltonian. In the following, we aim to provide additional insights into the dynamics in these three regimes by analyzing their respective ergodicity and thermalization properties. In particular, it will be the goal to further characterize the slow crossover regime by studying temporal fluctuations and correlations.

One of the key properties of systems obeying the Eigenstate Thermalization Hypothesis (ETH) is that long-time temporal fluctuations of expectation values of observables after a quench are exponentially small in the system size~\cite{srednicki_99, rigol_08, dalessio_15}. Equivalently, in the long-time limit the density matrix, from the point of view of local observables, is exponentially close to its time average at almost all times.  Moreover, this exponential scaling can serve as a defining criterion to check whether the observables are equilibrated, especially when the exact Hamiltonian is not accessible and one cannot analyze the level statistics. Hence, this represents a well-suited criterion that can be utilised to investigate thermalisation both experimentally and numerically.

Let us define the stroboscopic temporal fluctuations $\delta \mathcal{O}$ of an expectation value of an operator $\mathcal{O}$: $\mathcal O_\psi = \langle\psi|\mathcal{O}|\psi\rangle$, as measured over $N_T$ periods:
\be
\overline{\delta \mathcal{O}}_\psi = \sqrt{\frac{1}{N_T} \sum_{l=1}^{N_T} \left[  \mathcal{O}_\psi(lT) - \overline{\mathcal{O}_\psi}  \right]^2 }.
\ee
In isolated ergodic systems, according to ETH, thermalization implies that for any physical observable $\overline{\delta \mathcal{O}}_{\psi} \approx e^{-S/2}$, where $S\propto L$ is the thermodynamic entropy of the system. This ETH prediction implies that from the point of view of observables the state $|\psi(t)\rangle$ at almost all times is equivalent to the time-averaged density matrix, up to terms exponentially suppressed in the system size. In Floquet systems it is hard to define a thermodynamic entropy as all the Floquet energies are defined modulo $\Omega$ and thus the density of Floquet energy states is uniform. This is in agreement with expectations from thermodynamics that any thermal state of a Floquet system corresponds to infinite temperature, and is thus characterized by a flat density of states.  On the other hand, in the high-frequency driving regime for the finite systems we consider here the system does not heat up, and one can intuitively expect that one should use the entropy of an approximate extensive Floquet Hamiltonian, which can be computed perturbatively within a high-frequency expansion~\cite{rahav_03_pra,goldman_14,bukov_14,eckardt_15,mikami_15}. Alternatively, one can use the fact that in ergodic systems $S\approx S^F_{\psi,d}$, where $S^F_{\psi,d}$  is the diagonal entropy [von Neumann entropy of the time-averaged density matrix]~\cite{dalessio_15}. The diagonal entropy is readily computable from projecting the wave function of the system onto the exact Floquet eigenstates and does not depend on folding the spectrum. Then one can use this value of $S^F_{\psi,d}$ to estimate the expected scaling of $\overline{\delta \mathcal{O}}_\psi$ and compare with the numerical results.

The main plot in Fig.~\ref{fig:fluctuations} shows how the fluctuations of $\mathcal{O} = H_F^{(0)}$ decay with the system size. We compare the long-time average obtained with the Lanczos algorithm (red) to the infinite-time limit from the diagonal ensemble (blue). In both the high and the low-frequency regimes this decay is consistent with exponential, with the exponent close to the one expected from ETH (green dashed lines), and hence the system thermalises. Clearly, slight deviations are visible which, however, might result from finite-size effects as we are not able to extrapolate to the thermodynamic limit. To fully clarify this, it would be necessary to study larger system sizes which, however, is not possible within the used methodology. This thermalization corresponds to a finite temperature in the high-frequency regime, set by the energy density $\overline{\mathcal E_\psi}$, and to infinite temperature in the low-frequency regime. Note that in an extended region near the crossover scale $\Omega^*$ the situation is fundamentally different, similarly to the slow evolution discussed above. Specifically, the fluctuations $\overline{\delta \mathcal{E}}_\psi$ are strong and irregular such that ETH is not fulfilled and the evolution is non-thermalizing (non-ergodic) in this regime. The inset of Fig.~\ref{fig:fluctuations} shows the infinite-time energy fluctuations, calculated with ED, versus the driving frequency for three different system sizes, indicating the frequency domain of strong temporal fluctuations. Because of the very slow dynamics, it has not been possible to determine the infinite-time properties on the basis of the Lanczos algorithm. Instead, we have used full ED here, which limits the system sizes up to $L=16$. In the inset of Fig.~\ref{fig:fluctuations} one can see that the regime of strong temporal fluctuations of the energy with nonvanishing support over an extended frequency range features relatively sharp boundaries to the thermalizing regions. Upon increasing the system size, we observe a slight drift of this extended region to larger driving frequencies. On the basis of the system sizes accessible within our numerics, it is, however, unclear whether this region remains extended in the thermodynamic limit. Still, the extent over a few hopping amplitudes $J_0$ is substantial even for $L=16$ without a very strong finite-size dependence.

\begin{figure}
	\includegraphics[width=\columnwidth]{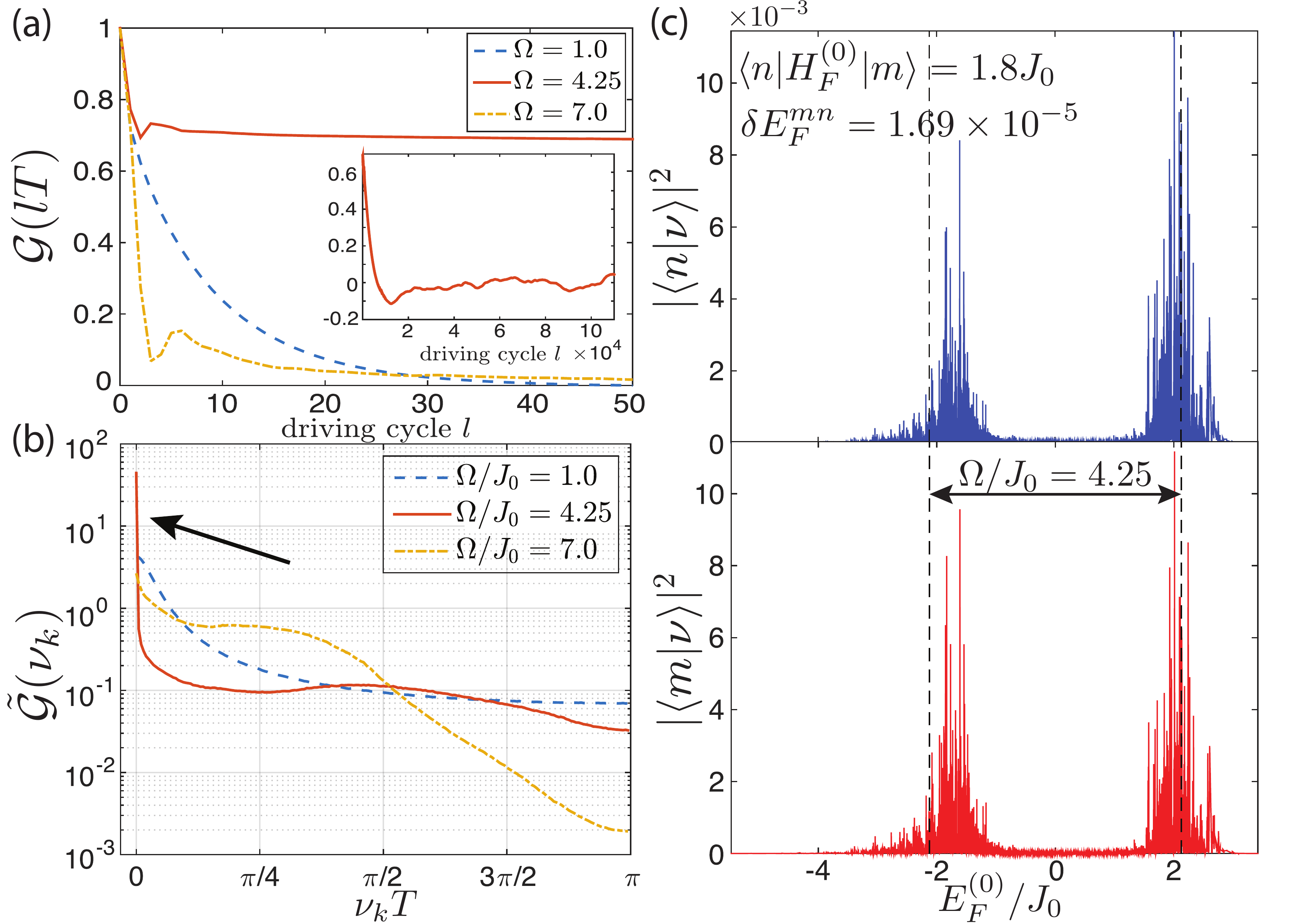}
	\caption{\label{fig:MB_resonances_main} The energy autocorrelator $\mathcal{G}$ as a function of time (a), and its Fourier transform $\tilde{\mathcal{G}}$ (b) for $\delta\nu T=\pi/200$. The arrow shows the many-body resonances peak. (c) A pair of many-body resonant Floquet eigenstates,
	$|m\rangle$ and $|n\rangle$, in the crossover regime. Here $|\nu\rangle$ are the eigenstates of the approximate Floquet Hamiltonian $H_F^{(0)}$ with energy $E_{F,\nu}^{(0)}$. The vertical dashed lines mark the boundaries of the Floquet zones, while the $x$-axis range corresponds to the many-body bandwidth. A similar procedure is used in time-of-flight images of superfluid Bose gases in optical lattices where quasimomentum states are projected onto momentum states to visualise the momentum (Bragg) peaks in nearby Brillouin zones. The parameters are $U/J_0=1$, $\zeta = 0.6$, $\delta\zeta = 0.12$, and $L=16$. }
\end{figure}

From the preceeding analysis we have seen that temporal fluctuations can become strong in the crossover region. In the following, we provide further evidence for nonergodic dynamics by studying temporal correlations. Specifically, an important indicator of non-thermalising evolution -- the long memory of fluctuations -- becomes manifest in the anomalously slow decay of nonequal-time correlation functions. To study this we now focus on the energy autocorrelation function:
\begin{eqnarray}
\mathcal{G}(lT) &=&  \frac{1}{\delta H_F^2}\sum_n  \langle n|  H_F^{(0)}(lT) H_F^{(0)}(0) |n\rangle_c\nonumber\\
&=& \frac{1}{\delta H_F^2}\sum_{m\neq n} |\langle n|H_F^{(0)}|m\rangle|^2 e^{-i(E_F^{m} - E_F^{n})lT},
\label{eq:E_autocorrelator_time}
\end{eqnarray}
where $|n\rangle$ is an eigenstate of the exact Floquet operator $U_F$ corresponding to the eigenvalue $\exp [-iE_F^n T]$. In the definition of $\mathcal{G}(lT)$ we have included the average variance $\delta H_F^2 = \sum_n |\langle n| [H_F^{(0)} - \langle H_F^{(0)}\rangle]^2|n\rangle|$ for normalization such that $\mathcal{G}(0)=1$. We sum over all eigenstates of $U_F$ to obtain better statistics. Consequently, $\mathcal{G}(lT)$ measures temporal correlations over the full many-body spectrum which goes beyond what we have studied before, where we have determined the dynamics starting from the ground-state manifold. The dynamics of $\mathcal{G}(lT)$, obtained from ED, we show in    Fig.~\ref{fig:MB_resonances_main}a. Although, in the absence of exact degeneracies,
for any finite system $\mathcal{G}(lT)\to 0$ as $l\to\infty$, the time scales which govern this decay differ tremendously between the thermalizing and the nonergodic regimes. Similar to the strong temporal fluctuations in the crossover region, we thus also find a very slow decay of temporal correlations which further supports the evidence for a strongly nonergodic regime separating the stable from the unstable phase.

\section{\label{sec:MB_res} Floquet Many-Body Resonances}

As we discussed in the previous Sections, the crossover regime exhibits nonergodic properties in terms of strong temporal fluctuations and correlations. In the following, we argue that this numerical observation can be related microscopically to the appearance of rare Floquet many-body resonances. To demonstrate this, we introduce the discretized Fourier transform of the energy autocorrelation function
\begin{equation}
\tilde{\mathcal{G}}(\nu_k)\!=\!\frac{1}{\delta H_F^2}\!\sum_{\substack{m\neq n }} |\langle n|H_F^{(0)}|m\rangle|^2\delta(\nu_k \leq |E_F^{n}-E_F^m|\leq \nu_{k+1})
\nonumber
\end{equation}
with $\nu_k=k\delta \nu$, $k\in\mathbb{N}$, and $\delta\nu$ a small quasienergy shell, see Fig.~\ref{fig:MB_resonances_main}b. Interestingly, in the crossover regime,  it features a well-pronounced peak near zero frequency, implying that near-resonant pairs of states of very small quasienergy difference dominate the long-time physics.  In terms of their physical energy, these pairs of states differ by integer multiples of the driving frequency and, therefore, represent resonances in the many-body spectrum.  It has been argued that these resonances lead to the breakdown of adiabatic perturbation theory in periodically-driven systems~\cite{russomanno_15,pweinberg_15}, and their manifestation in the form of non-analyticity in expectation values of observables has been studied for integrable systems~\cite{russomanno_12}.

By looking closer at the spectral properties, we can finally give an explanation for the observed heating--no-heating crossover. While we find many-body resonances over the full range of driving frequencies, their influence onto the dynamics differs substantially in the three observed regimes. In the high-frequency limit $J_0,U\ll \Omega$, the resonances are so weak and rare that they do not affect the dynamics of the system. The absorption of one quantum of $\Omega$ at these elevated energies requires the excitation of a complex many-body state due to the locality of $H_F$ in this regime -- a process which is at least exponentially suppressed in frequency. Entering the crossover regime $\Omega\approx\Omega^*$, a small amount of the resonant pairs begin to exhibit a very strong coupling, such that there is always some small number of eigenstates of the Floquet operator which carry significant weights in nearby Floquet zones, see
Fig.~\ref{fig:MB_resonances_main}c. This results in large matrix elements on the order of a few $J_0$, which represent a small but significant fraction of the total off-diagonal matrix elements of $H_F^{(0)}$ and which, according to Eq.~\eqref{eq:E_autocorrelator_time}, determine the slow dynamics of the system. These rare resonances cannot be neglected anymore but rather dominate the long-time dynamics leading to a very slow non-thermalising time evolution. This observation, that rare resonances dominate the low-energy spectral properties, is reminiscent of Griffith phases in disordered systems, but here for a system without disorder. Once the driving frequency is lowered further, $\Omega\ll\Omega^*$, the many-body resonances proliferate and the eigenstates of the Floquet operator become quite delocalized over the Floquet zones in the eigenbasis of $H_F^{(0)}$, see Sec.!\ref{sec:FMB_resolution}.  At the same time, the distribution of the off-diagonal matrix elements of $H_F^{(0)}$ becomes more uniform [see blue curve in Fig.~\ref{fig:MB_resonances_main}b]. This delocalization of the Floquet eigenstates in energy signifies rapid transfer of energy between the system and the drive, and the system quickly heats up to infinite temperature.

\section{\label{sec:FMB_resolution} Resolving the Resonances with the Inverse-Frequency Expansion}
	
The Floquet many-body resonances defined in the previous section were identified by projecting the exact Floquet eigenstates to the eigenstates of the infinte-frequency Floquet Hamiltonian $H_F^{(0)}$, which represents the leading order of the high-frequency expansion for the Floquet Hamiltonian. One can anticipate that these resonances can be made narrower and better defined if instead of $H_F^{(0)}$ one uses on a better approximate local Floquet Hamiltonian $\tilde H_F$, which can be e.g.~obtained by including higher-order terms in the inverse-frequency expansion. Indeed, physically the Floquet resonances occur when the rate of absorption and emission of photons from and to the drive is much smaller than the drive frequency. Without such resonances the system is described by $\tilde H_F$, whose exponential is a close approximation to the exact Floquet operator, i.e.~$U_F \approx \exp{(-i\tilde H_F T)}$, but whose spectrum is extensive.

At high frequencies, the exact eigenstates $|n\rangle$ of $U_F$ can all be assigned energies and each have high overlap with corresponding eigenstates of $\tilde H_F$. The Floquet many-body resonances occur at frequencies where this assignment is beginning to break down: they represent eigenstates of $U_F$ that appear as linear combinations of two (or more) eigenstates of $\tilde H_F$ that differ in energy by almost exactly one (or more) photon. In the regime we are considering, the eigenstates of $\tilde H_F$ that are involved in the resonances are typical thermalising states [in the sense of the Eigenstate Thermalisation Hypothesis] with nonzero entropy density, so each resonant state involves many ``bare'' configurations of the system; this is why we call them ``many-body'' resonances. In contrast, in noninteracting tight-binding systems, drive-assisted resonances can occur only when the frequency is smaller than the single-particle bandwidth of the Floquet Hamiltonian, which remains bounded in the thermodynamic limit.
	
\begin{figure*}[h!]
		\includegraphics[width=1.75\columnwidth]{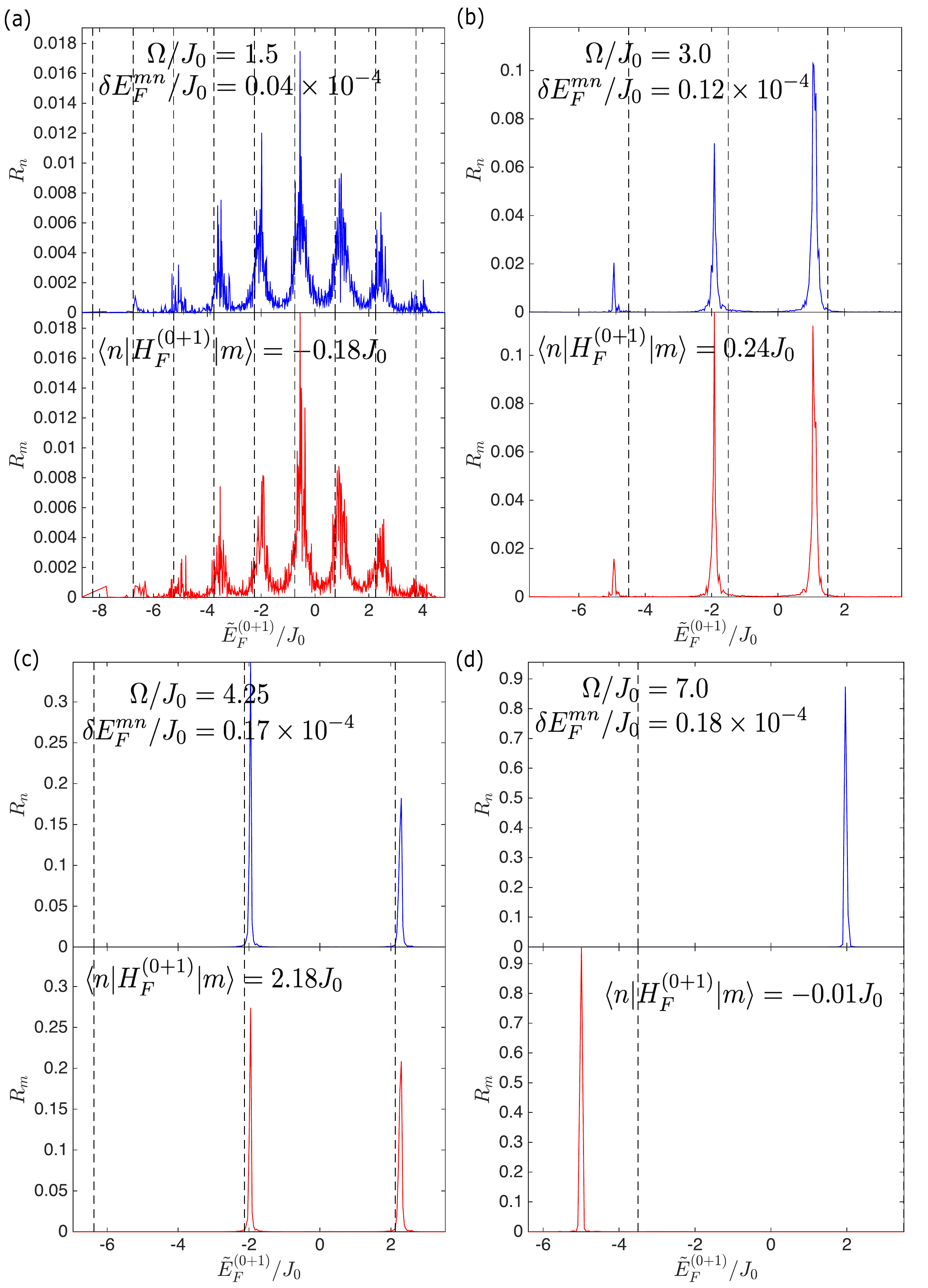}
		\caption{\label{fig:MB_resonances_app} (Color online).  Examples of nearly-degenerate pairs of exact Floquet eigenstates, including Floquet many-body resonances, in the dynamical regimes of interest. The resonant state $|n\rangle$ is quantified by the quantity $R_n= \sum_{\mu} |\langle n|\mu\rangle|^2\delta(\nu_k \leq \tilde E_F^{(0+1),\mu} \leq \nu_{k+1})$, with $\nu_k=k\delta \nu$, $k\in\mathbb{N}$, and the small energy shell $\delta\nu = \Omega/100$. Here $|\mu\rangle$ denotes an eigenstate of $H_F^{(0+1)}$. The range of the $x$-axis coincides with the many-body bandwidth, while the vertical dashed lines mark the boundaries of the Floquet zones.  The parameters are $U/J_0=1$, $\zeta = 0.6$, $\delta\zeta = 0.12$, which amounts to $J'/J_0 = 0.41$, $J/J_0=0.29$, and $L=16$.}
	\end{figure*}

	Floquet many-body resonances are beyond the van Vleck inverse-frequency expansion.  While we do not show evidence for this here, there are strong indications that the inverse-frequency expansion does not capture the hybridization of Floquet eigenstates in different Floquet zones due to photon absorption and, consequently, it also misses the appearance of many-body resonances~\cite{pweinberg_15}.  This can be understood intuitively from the fact that the inverse-frequency expansion necessarily produces an unfolded Floquet spectrum to every order. In fact the very requirement that the approximate Floquet Hamiltonian $\tilde H_F$ is local and extensive guarantees that the folded spectrum of the Floquet operator $U_F$ will have an extensive (per each eigenstate) number of unavoided level crossings corresponding to the photon resonances. Nevertheless, we shall now show that the Floquet many-body resonances can be nicely resolved using the approximate Floquet Hamiltonian, including the leading order correction.  To this end, we proceed as follows:
	\begin{itemize}
		\item[(i)] We first calculate an approximation to the Floquet Hamiltonian using the van Vleck high-frequency expansion $H_F^{(0+1)}$. In the present discussion we stop after we take into account the leading $\Omega^{-1}$--correction, see App.~\ref{app:Floquet corrections}. It is interesting to note how much resolution one gains by including only the first $\Omega^{-1}$-correction [compare Fig.~\ref{fig:MB_resonances_main}(c) and Fig.~\ref{fig:MB_resonances_app}(c) below which show the same resonant pair resolved with the zeroth and first correction, respectively].
		\item[(ii)] Diagonalise $H_F^{(0+1)}$; denote its eigenenergies by $E_F^{(0+1)}$ and its eigenstates by $|\nu\rangle$.
	\end{itemize}
	In principle, to visualise a Floquet many-body resonance it suffices to project a candidate eigenstate $|n\rangle$ of $U_F$ onto the eigenstates $|\nu\rangle$ of $\tilde H_F$, and map out a probability distribution as a function of the energy $\tilde E_F$.  This reveals the Floquet zones in which the resonant states have most of their weight. It works because the inverse-frequency expansion necessarily produces an unfolded Floquet spectrum, as it becomes exact at infinite-frequencies.  This procedure is analogous to time-of-flight imaging in cold atom systems, where one projects a Bose-Einstein condensate formed in an optical lattice onto free space, and reads off the quasimomentum peaks and their weights from the interference image.  Figure~\ref{fig:MB_resonances_main}c above is obtained after applying points (i) and (ii) to the Hamiltonian $H_F^{(0)}$.
	
	The above two points are indeed enough to show the existence of many-body resonances, localised in neighbouring Floquet zones.  However, by looking at the distance between the resonance peaks, we find that the approximation [e.g.~$H_F^{(0+1)}$] to $\tilde H_F$ obtained from the
	inverse-frequency expansion does not ``know'' the correct value of $\Omega$. Thus,  the resonant peaks after applying (i) and (ii) differ in energy by more than $\Omega$.  Therefore, we choose to correct the eigenenergies $E_F^{(0+1)}$ as follows:
	\begin{itemize}
		\item[(iii)] We calculate the expectation value of the exact Floquet operator in the approximate eigenstates, $\langle\nu|U_F|\nu\rangle$. In the regime of resonances, this gives complex numbers of magnitude close to unity.  Hence, we obtain quasienergies for each state as $\mathcal{K}_{F,\nu}^{(0+1)} = i/T\log\left[\langle\nu|U_F|\nu\rangle/|\langle\nu|U_F|\nu\rangle|\right]$.
		\item[(iv)] Last, one has to unfold the spectrum to get the ``revised'' energies $\tilde E_{F,\nu}^{(0+1)}$. For this purpose, one can plot $\mathcal{K}_{F,\nu}^{(0+1)}$ vs.~$E_{F,\nu}^{(0+1)}$ for each state.  At high enough frequency these points are all near smooth curves with slope near one in each Floquet zone, thus providing a natural unfolding of the spectrum.  But with this unfolding the energies do not properly match the quasi-energies.  To get the proper revised energies $\tilde E_{F,\nu}^{(0+1)}$ we do two more steps:  First, we shift all energies $E_{F,\nu}^{(0+1)}$ by some smooth function (in practice a linear function suffices) of the energy, to make the spectrum all close to $E_{F,\nu}^{(0+1)}\approx\mathcal{K}_{F,\nu}^{(0+1)}  \ \mathrm{mod} \ \Omega$.  Thus in the linear approximation we define a revised approximate Hamiltonian as $\tilde H = b + m H_F^{(0+1)}$, with $m$ near one and a shift $b$ of the zero of energy.  Then, finally, we add a small amount to each energy to make the revised energies $\tilde E_{F,\nu}^{(0+1)}$ precisely match the quasi-energies $\mathcal{K}_{F,\nu}^{(0+1)}$, modulo $\Omega$.  Thus we have produced a revised approximate Floquet Hamiltonian $\tilde H_F$ whose eigenstates are identical to those of $H_F^{(0+1)}$, but whose spectrum has been shifted to agree with the $\mathcal{K}_{F,\nu}^{(0+1)}$.
	\end{itemize}
	Step (iv) of this procedure fails at low frequency, where many states have $|\langle\nu|U_F|\nu\rangle|\ll 1$ and thus do not have well-defined quasi-energies.  This results in ambiguities in the unfolding procedure (iv). For the model under consideration, we have found that for $L \leq 16$ we obtain meaningful and reliable revised energies for $\Omega/J_0\gtrsim 1.5$. Interestingly, this frequency is significantly less than the crossover scale $\Omega^\ast$ suggesting that the heating transition occurs through proliferation of these resonances in the regime where they are still narrow and well defined.
	
	Figure~\ref{fig:MB_resonances_app} shows four nearly-degenerate pairs of exact Floquet eigenstates at different values of $\Omega/J_0$.  To take into account the effect of the density of states, we sum the projections $|\langle n|\nu\rangle|^2$ over a small shell of revised approximate energies, see caption.
	At high-frequencies, Fig.~\ref{fig:MB_resonances_app}(d), we do not find resonances.  Here the matrix elements $\langle \nu|U_F|\mu\rangle$ between states in different Floquet zones are all small compared to the quasi-energy level spacing in the spectrum of $U_F$, so even almost-degenerate eigenstates of $U_F$ map almost purely on to a single Floquet zone.  Thus in this regime the spectrum of $U_F$ can be unambiguously unfolded, and an excellent local approximation to the exact $H_F$ exists.  Whether or not this regime inevitably ``retreats'' to infinite $\Omega$ as $L\rightarrow \infty$ is an interesting question for future investigation. As the frequency is decreased, Fig.~\ref{fig:MB_resonances_app}(c) and Fig.~\ref{fig:MB_resonances_app}(b), Floquet many-body resonances do appear.  We find that the matrix element between resonant states $\langle m|H_F^{(0+1)}|n\rangle$ is enhanced up to a few times the bare hopping amplitude $J_0$.  As a result, for intermediate frequencies, these Floquet many-body resonances constitute the dominant fraction of off-diagonal matrix elements of the energy operator $H_F^{(0+1)}$.  Moreover, they connect different Floquet zones, and the system thus starts absorbing (or emitting) energy.  Consequently, due to the small number of resonant pairs with large off-diagonal matrix elements, the dynamics of the energy is, to a large extent, dominated by these drive-induced transitions, which leads to the observed non-thermalizing glassy behaviour.  It follows that a description based on statistical mechanics w.r.t.~the approximate Hamiltonian $H_F^{(0+1)}$ fails to capture the stroboscopic physics at any sensible time scale in this crossover regime.  In this same crossover regime, we also find that the eigenstates of $H_F^{(0+1)}$ can be cleanly assigned quasi-energies, so there is a well-defined ``folding'' procedure, see steps (iii) and (iv) above, to define the energies $\tilde E_{F,\nu}^{(0+1)}$,  but the unfolding of the \emph{exact} quasispectrum of $U_F$ is no longer well-defined, due to the presence of the Floquet many-body resonances.  Finally, Fig.~\ref{fig:MB_resonances_app}(a), when the driving frequency is reduced even further, the Floquet many-body resonances proliferate.  At the same time, however, the matrix elements $\langle m|H_F^{(0+1)}|n\rangle$ between the resonant states decrease again and become closer to the average off-diagonal matrix element [which is small since these states are well-thermalized to infinite temperature]. Hence, the system continuously absorbs energy and heats up to infinite temperature, thereby delocalising along the energy ladder.  This heating is rapid, as indicated by the broad linewidths in Fig.~\ref{fig:MB_resonances_app}(a).  The dynamics of the system is completely chaotic and, therefore, thermalizing again.  Decreasing the frequency even further to $\Omega/J_0 = 1$, $\tilde H_F$ is no longer well-defined, as we explained above, while $H_F^{(0+1)}$ is becoming a very poor approximation to the correct, now highly-nonlocal $H_F$.  Hence, the eigenstates of $U_F$ are completely delocalised over the $E_F^{(0+1)}$--axis.
	
Applying the van Vleck inverse-frequency expansion to a given order $n_\mathrm{HFE}$ yields a truncated (approximate) Floquet Hamiltonian $H_F^{(0+\dots+n_\mathrm{HFE})}$ and the corresponding truncated (approximate) time-periodic Kick operator $K_F^{(0+\dots+n_\mathrm{HFE})}(t)$. The Hamiltonian $H_F^{(0+\dots+n_\mathrm{HFE})}$ is a sum of local many-body operators with an unfolded spectrum, the bandwidth of which necessarily goes to infinity in the thermodynamic limit. If we now use this truncated kick operator to transform the original lab-frame Hamiltonian $H(t)$ to a rotating frame, the corresponding rot-frame Hamiltonian has the form $\tilde H^\mathrm{rot}(t) = H_F^{(0+\dots+n_\mathrm{HFE})} + W(t)$, where $W(t)=W(t+T)\sim\Omega^{-(n_\mathrm{HFE}+1)}$ by construction~\cite{rahav_03_pra,goldman_14,abanin_15_2}. In this rotating frame, we can interpret the heating problem as follows: the inverse-frequency expansion takes care only of the virtual photon-absorption processes, pretty much like any ordinary Schrierffer-Wolff transformation does~\cite{bukov_15_SW}. As a result, this shifts the energy levels of the non-driven Hamiltonian $H_0$ by a small amount. This is why the width of the resonances is reduced tremendously by taking into account the leading-order correction, compare Fig.~\ref{fig:MB_resonances_main}(c) and Fig.~\ref{fig:MB_resonances_app}(c). Although these virtual transitions do have an effect on the underlying physics, they can only result in heating to a small finite temperature [e.g.~due to the abruptly switching on the drive or a possible adiabatic preparation of the initial state]. At this level, if one insists that the spectrum of the Hamiltonian $H_F^{(0+\dots+n_\mathrm{HFE})}$ is only defined modulo $\Omega$ and folds it artificially, the original Wigner-Dyson level spacing statistics of the non-integrable $H_F^{(0+\dots+n_\mathrm{HFE})}$ will suddenly change to Poisson statistics, due to the lack of photon-assisted level repulsion, see App.~\ref{app:level_statistics}. On the other hand, taking back into consideration the time-dependent piece $W(t)$, we find that it is responsible for driving real photon-absorption transitions between the approximate Floquet levels of $H_F^{(0+\dots+n_\mathrm{HFE})}$, which are not captured by the inverse-frequency expansion to any order. Note that these pairs of states with energy difference $E_{F,m}^{(0+\dots+n_\mathrm{HFE})} - E_{F,n}^{(0+\dots+n_\mathrm{HFE})}\approx l\Omega$ with $l\in\mathbb{N}$ are guaranteed to exist in the TD limit where the spectrum becomes dense and unbounded. It is these direct transitions between the Floquet many-body states of $H_F^{(0+\dots+n_\mathrm{HFE})}$ which can potentially lead to heating to infinite temperature in the longer run, irrespective of the driving frequency. Ultimately whether this heating happens or not in the thermodynamic limit will be determined by the ratio of the width of the many-body resonances in the basis of $H_F^{(0+\dots+n_\mathrm{HFE})}$ and the splitting between these resonances due to $W(t)$. We leave this interesting and important question for future work.

%%%%%%%%%%%%%%%%%%%%%%%%%%%%%
%	Discussion

\section{\label{sec:conclusion} Outlook and Discussion}

In summary, we presented numerical evidence that strongly interacting two-band systems which are resonantly coupled via a periodic drive feature a large window of stable controllable time-evolution at high frequencies. The studied two-band system only weakly absorbs energy from the drive at the experimentally-relevant time scales and is, therefore, amenable to Floquet engineering. This opens up the possibility of studying also other interesting strongly interacting systems including, for example, fractional Floquet topological insulators~\cite{grushin_14} or Heisenberg models with artificial gauge fields~\cite{bukov_15_SW}. By studying the heating--to--no-heating crossover, we laid the foundations to understand the microscopic origin of heating in non-integrable perodically-driven systems.

It is important to emphasize, that our two-band model might not be fully sufficient to describe all experiments, due to the presence of even higher bands. However, their influence on heating, can be estimated from our results. Although the typical driving frequencies may not necessarily be large enough to induce direct transitions to these bands, higher-order photon absorption processes with reduced matrix elements can occur~\cite{weinberg_15}. Since higher bands have much larger bandwidths, it becomes much more likely to hit a single-particle resonance which defines the crossover scale $\Omega^\ast$, cf.~Sec.~\ref{subsec:U_dep}. If such a single-particle resonance is present, we expect that we will again see heating.  Last, while we did not consider this, it also bears mentioning that the presence of perpendicular to the lattice plane dimensions, comprising continuous degrees of freedom (tubes/pancakes), plays a crucial role for heating.  In such cases, heating effects are enhanced by photon-stimulated scattering into these additional dimensions, which can act as reservoirs and facilitate thermalisation at a higher temperature
~\cite{bilitewski_14,bilitewski_15,choundhury_15,choundhury_15_2}.

The existence of nonthermalizing time-evolution, featuring strong temporal fluctuations and correlations, at the crossover between a stable and an unstable regime is reminiscent of a dynamical phase transition between many-body localised and delocalised phases in energy space~\cite{dalessio_13}.  We have identified many-body resonances as the microscopic origin of this behaviour.  Nevertheless, our results do not allow for a direct extrapolation to the thermodynamic limit.  Whether or not infinite isolated ergodic Floquet systems at high-frequencies eventually heat up to infinite temperature at infinite times or remain localised in energy space forever, remains yet to be revealed.  While this is still an open problem with examples existing indicative of either outcome~\cite{prosen_98a,prosen_98b,prosen_99,prosen_02,dalessio_13,dalessio_14,lazarides_14,lazarides_14_2,ponte_15,das_10,roy_15,genske_15,russomanno_15_epl,citro_15,straeter_16}, recently developed rigorous proofs suggest that heating in fermionic and spin systems, if at all present, happens at most exponentially slowly in the driving frequency~\cite{abanin_15,kuwahara_15,mori_15,abanin_15_2}.

\begin{acknowledgments}
We thank D.~Abanin, Ye.~Bar Lev, I.~Bloch, L.~D'Alessio, M.~Dolfi, S. Gopalakrishnan, T. Grover, V. Khemani, M.~Kolodrubetz, A.~Lazarides, U.~Schneider and C. Weitenberg for insightful and interesting discussions, and especially acknowledge the help of P.~Weinberg for co-developing the exact diagonalisation code used in this work.  D. H. is the Addie and Harold Broitman Member at I.A.S.  This work was supported by AFOSR FA9550-13-1-0039, NSF DMR-1506340, ARO W911NF1410540, and the Deutsche Akademie der Naturforscher Leopoldina (grant No. LPDS 2013-07 and LPDR 2015-01). The computational work reported on in this paper was partly performed on the Shared Computing Cluster which is administered by Boston University's Research Computing Services. URL: www.bu.edu/tech/support/research/. The authors also acknowledge the Research Computing Services group for providing consulting support.
\end{acknowledgments}

\bibliographystyle{apsrev4-1}
\bibliography{Floquet_bib}

\begin{appendix}
	
\begin{widetext}

	\section{\label{app:definitions_obs}  Microscopic Definitions for the Observables and Entropies Pertinent to Heating.}

	In this section we will define all key observables and entropies analyzed throughout the paper. Let us denote by $\{|n\rangle\}$ the eigenstates of the exact many-body Floquet operator $U_F = \mathcal{T}_t\mathrm{exp}\left(-i\int_0^TH(t)\mathrm{d}t\right)$, and by $\{|\nu\rangle\}$ -- the eigenstates of the approximate Floquet Hamiltonian $H_F^{(0)}$ obtained in the leading order in the inverse-frequency expansion. Note that $H_F^{(0)}$ is a local Hamiltonian with unfolded spectrum so we can choose the initial state to be the ground state of $H_F^{(0)}$, which we denote by $|\psi\rangle$ such that $H_F^{(0)}|\psi\rangle = E_F^{(0)} |\psi\rangle$. We shall discuss how observables, defined below, can be extended to initial mixed states. The ``transition'' probability between an approximate and an exact Floquet eigenstate is given by $|\langle \nu |n \rangle|^2$. The transition matrix containing all these probabilities is denoted by $C_{\nu n} = C_{n\nu}=|\langle \nu |n \rangle|^2$.
	
	Assuming that there are no degeneracies in the exact Floquet spectrum, the stroboscopic diagonal expectation value of any observable $\mathcal O$ and its fluctuations are given by
	\begin{eqnarray}
	\langle \mathcal{O}\rangle_d &=& \lim_{N_T\to\infty}\frac{1}{N_T}\sum_{l=1}^{N_T}\langle\psi(lT)|\mathcal{O}|\psi(lT)\rangle = \sum_n\langle n|\mathcal{O}|n\rangle C_{n\psi},\nonumber\\
	\langle \delta\mathcal{O}\rangle_d &=& \sqrt{ \lim_{N_T\to\infty}\frac{1}{N_T}\sum_{l=1}^{N_T} \bigg( \langle\psi(lT)|\mathcal{O}|\psi(lT)\rangle - \langle \mathcal{O}\rangle_d \bigg)^2 } = \sqrt{  \sum_{n\neq m}|\langle n|\mathcal{O}|m\rangle|^2 C_{n\psi} C_{m\psi}  }
	\end{eqnarray}
	%As explained in the main text, to study heating in the lab frame, it suffices to consider stroboscopic evolution.
	In order to define how much energy is pumped into the system by the drive, we measure the energy associated with the approximate Floquet Hamiltonian, i.e~ we choose $\mathcal O=H_F^{(0)}$. The diagonal expectation value then becomes
	\begin{eqnarray}
	\langle\psi| H_F^{(0)}|\psi\rangle_d = \sum_n\langle n|H_F^{(0)}|n\rangle C_{n\psi} = \sum_{\nu} E_{F,\nu}^{(0)}p_{\nu\psi},
	\end{eqnarray}
	where $p_{\nu\psi} = \sum_n C_{\nu n}C_{n\psi}$ is the probability to occupy the $\nu$-th eigenstate of $H_F^{(0)}$ in the diagonal ensemble (i.e.~for $t\to\infty$), starting from its GS $|\psi\rangle$. The transition probability matrix $p$ can be also understood as a result of a double quench, where the system is prepared in the ground state of $H_F^{(0)}$. Then it is evolved periodically according to the Hamiltonian $H(t)$ and after many periods $N_T\to\infty$, it is projected back to the basis of $H_F^{(0)}$. It is easy to see that under these conditions the transition probability becomes a Markov matrix and satisfies the factorization property (see also Ref.~\cite{dalessio_15} for more details).
	
	%\begin{comment}
	We can now define the following infinite-time quantities, which are used to analyze heating in the system:
	\begin{itemize}
		\item Normalized energy (or equivalently normalized work) $\overline{Q}_\psi$ pumped into the system during the drive:
		\begin{equation}
		\label{eq:Q_inf_time}
		\overline{Q}_\psi = \frac{\langle\psi| H_F^{(0)}|\psi\rangle_d - E_{F,\psi}^{(0)}  }{ E_{F,\beta=0}^{(0)} -E_{F,\psi}^{(0)} },
		\end{equation}
		where $E_F^{(0)}=\langle \psi| H_F^{(0)}|\psi\rangle$ is the ground state energy of $H_F^{(0)}$, $E_{F,\beta=0}^{(0)} = 1/\mathcal D \sum_\nu E_{F,\nu}^{(0)}$ is the energy at infinite temperature and $\mathcal D$ is the dimensionality of the Hilbert space.  For the system considered in this paper, in the thermodynamic limit $L\to \infty$, $E_{F,\beta=0}^{(0)}/L\to 0$ [$E_{F,\beta=0}^{(0)}/L=-U/(4L)$ for half-filling].
		\item Normalized diagonal (double-quench) entropy $\mathcal{S}_\psi$:
		\begin{equation}
		\label{eq:Sd_inf_time}
		\mathcal{S}_\psi = \frac{ S_{\psi,d} - S_{\psi}^{(0)}  }{ S_{\beta=0} - S_{\psi}^{(0)} } = \frac{ S_{\psi,d}   }{ S_{\beta=0}  },
		\end{equation}
		where $S_{\psi,d} = -\sum_{\nu} p_{\nu\psi}\log p_{\nu\psi}$ is the entropy in the diagonal ensemble in the basis of $H_F^{(0)}$, i.e.~with $p_{\nu\psi} = \sum_n C_{\nu n}C_{n\psi}$.  The initial state is the ground state of $H_F^{(0)}$ and therefore $S_{\psi}^{(0)} = 0$, while the maximum possible entropy (at infinite-temperature) is $S_{\beta=0} = L\log 2$.  This entropy characterizes the spreading of the initial state $|\psi\rangle$ over other eigenstates of $H_F^{(0)}$ after the system is driven for infinitely many periods. Note that there is a universal non-extensive correction to the entropy $S_{\psi,d}$ given by $\gamma-1$, where $\gamma$ is the Euler constant~\cite{ikeda_15}. This correction originates from the fact that the entropy is a non-linear function of the density matrix.
		\item Floquet diagonal entropy:
		\begin{eqnarray}
		S_{\psi,d}^F = -\sum_{n} C_{\psi n}\log C_{n\psi}.
		\end{eqnarray}
		This entropy measures spreading of the initial state $|\psi\rangle$ over the eigenstates of the Floquet Hamiltonian. It is equivalent to the von-Neumann's entropy of the (stroboscopically) time averaged density matrix of a driven system.			
		\item Normalized entanglement entropy of the half chain $\overline{\mathcal{S}^\mathrm{ent}_\psi}$ produced by the drive:
		\begin{eqnarray}
		\overline{\mathcal{S}^\mathrm{ent}_\psi} &=& \frac{\overline{s^\mathrm{ent}_\psi} - s^\mathrm{ent}_\psi(t=0)}{\log(2) - s^\mathrm{ent}_\psi(t=0)},\nonumber\\
		\overline{s^\mathrm{ent}_\psi} &=&  \lim_{N_T\to\infty}\frac{1}{N_T}\sum_{l=1}^{N_T}\frac{1}{L/2}\mathrm{Tr}_{B} \left[ -\rho_B(lT) \log \rho_B(lT) \right]
		\end{eqnarray}
		Here, $B$ denotes the set of the first $L/2$ lattice sites, $\rho_B(lT)$ -- the reduced density matrix of $B$ at time $t=lT$, and $s^\mathrm{ent}_\psi(t=0)$ is the entanglement entropy of the initial state.
		\item Energy density fluctuations $\overline{\delta \mathcal{E}}_\psi$:
		\begin{eqnarray}
		\label{eq:dE_inf_time}
		\overline{\delta\mathcal{E}}_\psi =  \frac{1}{L}\sqrt{  \lim_{N_T\to\infty}\frac{1}{N_T}\sum_{l=1}^{N_T}\left(  \langle\psi(lT)| H_F^{(0)} |\psi(lT)\rangle   -  \langle\psi| H_F^{(0)} |\psi\rangle_d   \right)^2   }.
		\end{eqnarray}
	\end{itemize}
	%\end{comment}	

	\section{\label{app:finite_size_scaling} System Size Dependence. Comparison between Exact Diagonalisation and Lanczos Time Evolution.}

	The discussion in this section carries a two-fold purpose: (i) to study the system size dependence of the observables considered in the main text, i.e.~the normalised energy, its fluctuations, the entanglement and diagonal entropy, and (ii) to compare the long-time Lanczos dynamics of these quantities with the infinite-time ED expectation values defined in the previous section. For all the data presented in this section, we initiate the evolution from the ground state of the infinite-frequency Floquet Hamiltonian $H_F^{(0)}$, while we evolve with the exact time-dependent Hamiltonian $H(t)$. All measurements are taken stroboscopically.
	
	\begin{figure*}[h!]
		\includegraphics[width=\columnwidth]{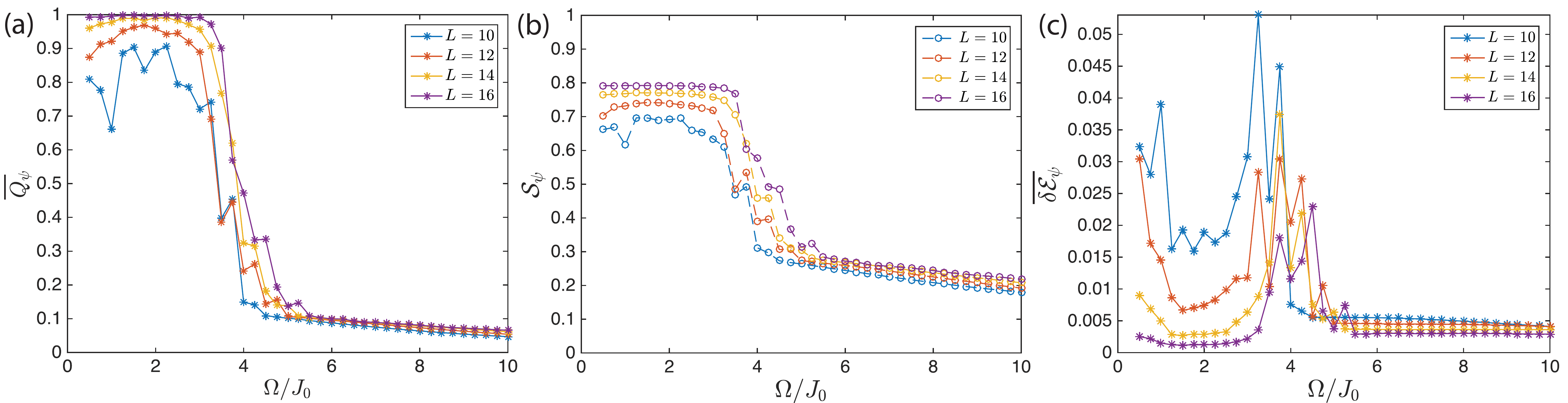}
		\caption{\label{fig:system_size_ED} (Color online). System size dependence of the exact diagonalisation results. (a) normalised energy, (b) diagonal entropy and (c) energy density fluctuations.  The parameters are $U/J_0=1$, $\zeta = 0.6$, $\delta\zeta = 0.12$, which amounts to $J'/J_0 = 0.41$, $J/J_0=0.29$. }
	\end{figure*}
	
	\emph{Exact Diagonalisation.} Exact diagonalisation (ED) allows us to discuss system sizes of up to $L=16$ sites, taking into account all symmetries present in the problem. Although these system sizes are admittedly far away from the realistic thermodynamic limit, ED is still a very useful tool, since it allows us to make statements about the infinite-time limit. Figure~\ref{fig:system_size_ED}\,(a) and (b) shows the infinite-time system-size dependence of the normalised energy and the relative diagonal entropy curves, respectively.  The data suggests a small drift of the transition region in the direction of increasing driving frequency. However, given that the drift is small and that the crossover frequency is close to the single-particle band-width based on this data we can not draw conclusions about the thermodynamic limit. Due to the presence of resonances in the crossover regime, we were unable to scale-collapse the data. Fig.~\ref{fig:system_size_ED}(c) shows the system size dependence of the energy density fluctuations. Clearly, the region of large fluctuations coincides nicely with the crossover between the infinite-heating and no-heating regimes.
	
	\begin{figure*}[h!]
		\includegraphics[width=\columnwidth]{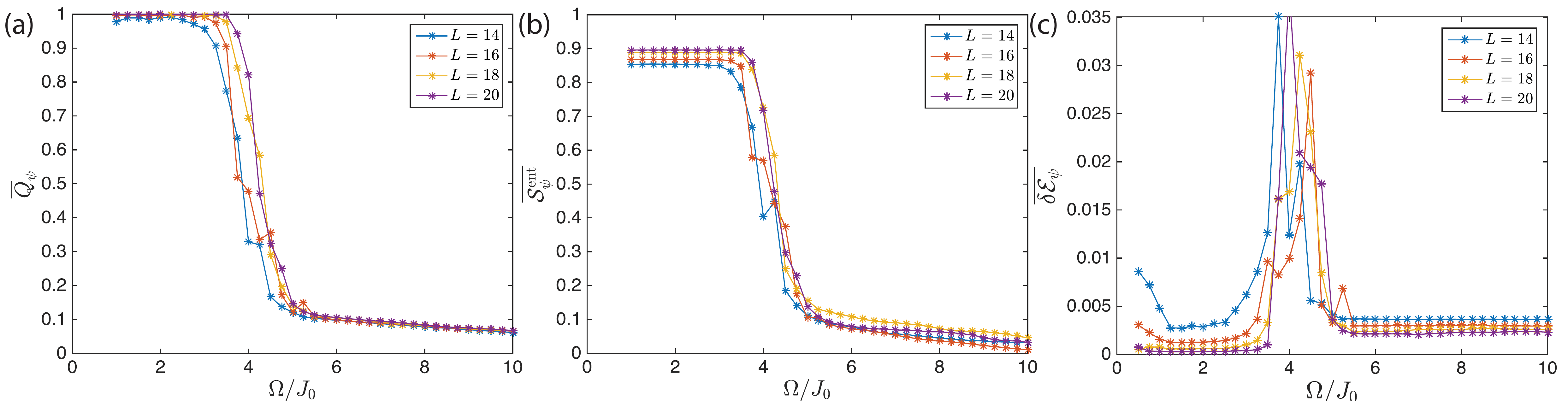}
		\caption{\label{fig:system_size_Lanczos} (Color online). System size dependence of the Lanczos evolution curves. (a) normalised energy, (b) entanglement entropy and (c) energy density fluctuations.  The parameters are $U/J_0=1$, $\zeta = 0.6$, $\delta\zeta = 0.12$, which amounts to $J'/J_0 = 0.41$, $J/J_0=0.29$. }
	\end{figure*}
	
	\emph{Lanczos Time Evolution.} For comparison, we also show the system-size dependence of the long-time averaged curves, obtained using Lanczos evolution. Figure~\ref{fig:system_size_Lanczos} (a), (b) and (c) show the system-size dependence of the normalised energy, the entanglement entropy and the energy density fluctuations. Here we can go to larger system sizes, while the evolution is limited to finite, but long times. We evolve the initial state for $5000$ periods and average the data between periods $T_1 = 1000$ and $T_2 = 5000$, to make sure we avoid any initial transients. From this figure we see that the drift of the crossover frequency with the system size becomes almost negligible as we reach $L=20$. In Fig.~\ref{fig:comparison_Lanczos_ED} we show the comparison between the data obtained by the Lancsoz and ED methods. We see that in the two thermalized phases of low and high frequencies the two methods agree to excellent precision. In the glassy crossover region, however, the disagreement is significant due to extremely slow dynamics, which does not saturate after $5000$ periods.

	\begin{figure*}[h!]
		\includegraphics[width=\columnwidth]{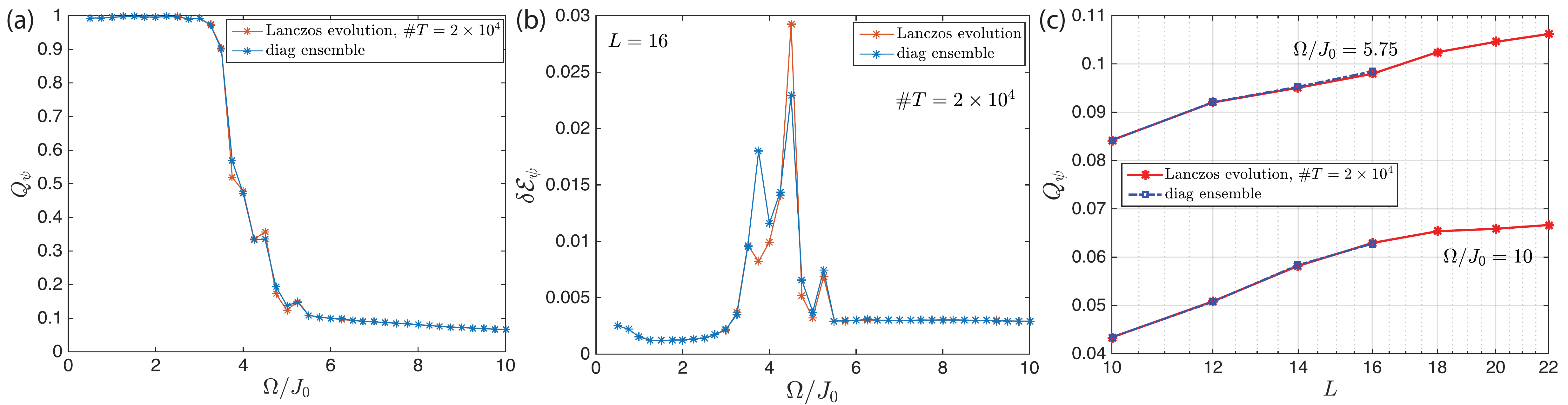}
		\caption{\label{fig:comparison_Lanczos_ED} (Color online). Comparison between infinite-time ED and long-time average of the exact Lanczos time evolution. Panels (a) and (c) show the normalised energy and energy-density fluctuations for $L=16$. In panel (a) we have assumed $E_{F,\beta = 0}^{(0)} = 0$. Panel (b) shows the system-size dependence of the normalised energy on a logarithmic scale. The parameters are $U/J_0=1$, $\zeta = 0.6$, $\delta\zeta = 0.12$, which amounts to $J'/J_0 = 0.41$, $J/J_0=0.29$. }
	\end{figure*}

	To shed more light on the localisation-delocalisation dilemma, we choose two points from the $Q_\psi(\Omega/J_0)$ curve in
	Fig.~\ref{fig:comparison_Lanczos_ED} (a), both in the high-frequency localised region, and monitor the behaviour of the normalised
	energy as a function of the system size $L$, see Fig.~\ref{fig:comparison_Lanczos_ED} (c). In this regime, we observe a nice agreement
	between the infinite-time ED curves and the time-averaged Lanczos evolution data taken over $2\times 10^4$ driving periods.
	An interesting feature is observed if we plot the system-size dependence logarithmically: both the frequency closer to the
	transition region and the one deep into the thermalising phase feature apparently sublogarithmic growth. Moreover, the
	$\Omega/J_0=10$ curve seems to even saturate at large system sizes. If this trend remains to infinite $L$, that would mean that there
	is a true finite-frequency transition between a localized and a delocalized phase in the thermodynamic limit.

	\section{\label{app:finite-Temperature} Finite-Temperature Effects}

	Until now we focused on the system prepared in the initial ground state of $H_F^{(0)}$. In this section we check the sensitivity of the results to the presence of a finite temperature. Specifically, we assume that the system is initially prepared in a state according to the equilibrium Boltzmann distribution with respect to the Hamiltonian $H_\text{F}^{(0)}$. Technically, we initialize the system in one of the eigenstates of $H_F^{(0)}$, $|\nu\rangle$, with the probability given by the Gibbs distribution $\rho_\nu \propto \exp[-\beta E_{F,\nu}^{(0)}]$. Then we calculate all observables such as $\mathcal{E}_\psi = \langle\psi |H_F^{(0)}|\psi\rangle$, $S_{\psi,d}$ and $\delta \mathcal{E}_\psi$ for this eigenstate. Finally, we take the average of the result over all available eigenstates. The observables computed in this way characterize the delocalization of individual eigenstates exclusively due to the driving, and disentangles it from the initial thermal broadening. For instance, in the infinite-frequency limit, where the eigenstates of the Floquet Hamiltonian coincide with the eigenstates of $H_F^{(0)}$ the (eigenstate) diagonal entropy computed in this way, will be zero at any temperature as each initial eigenstate remains fully localized in energy space. In particular, we extend the definitions of the observables and entropies in the following way:

	\begin{figure*}[h!]
		\includegraphics[width=\columnwidth]{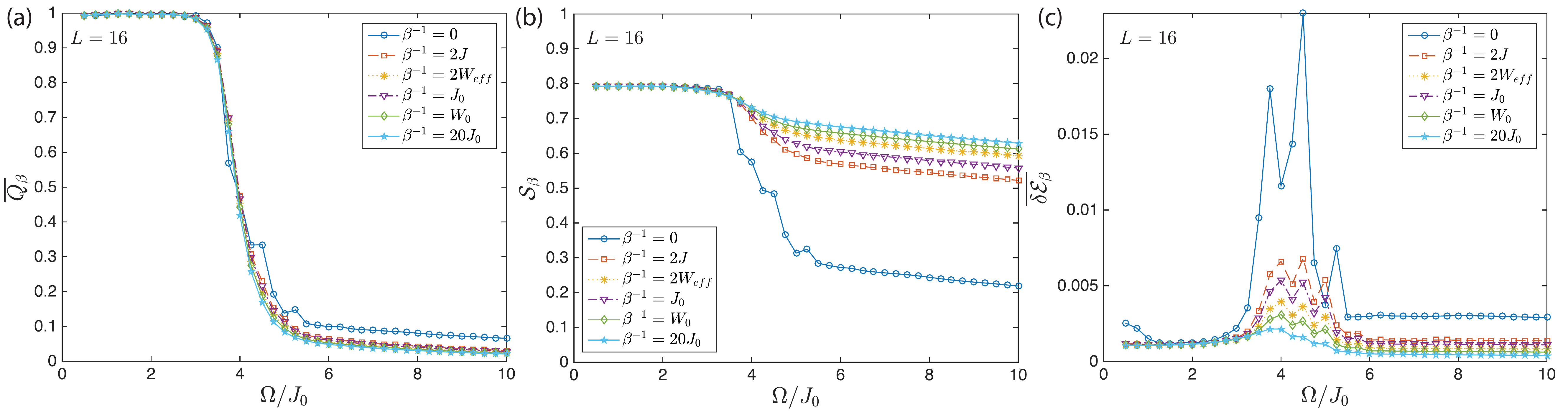}
		\caption{\label{fig:temperature} (Color online). Dependence of the infinite-time normalised energy $\overline{Q}_\beta$ (a), (eigenstate) diagonal entropy $\mathcal{S}_\beta$ (b), and energy-density fluctuations $\overline{\delta\mathcal{E}}_\beta$ (c) on the temperature $\beta^{-1}$ of the initial state for $L=16$. The parameters are $U/J_0=1$, $\zeta = 0.6$, $\delta\zeta = 0.12$, which amounts to $J'/J_0 = 0.41$, $J/J_0=0.29$. }
	\end{figure*}
	
	\begin{itemize}
		\item Dimensionless normalized energy $\overline{Q}_\beta$ starting from a finite-temperature state:
		\begin{equation}
		\overline{Q}_\beta = \frac{\sum_\nu \langle\nu| H_F^{(0)}|\nu\rangle_d\ \rho_\nu(\beta) - \sum_\nu E_{F,\nu}^{(0)}\rho_\nu(\beta) }{ E_{F,\beta=0}^{(0)}- \sum_\nu  E_{F,\nu}^{(0)}\rho_\nu(\beta) },
		\end{equation}
		\item Normalized (eigenstate) diagonal entropy $\mathcal{S}_\beta$ at finite-temperature:
		\begin{equation}
		\mathcal{S}_\beta = \frac{ \sum_{\nu}S_{\nu,d}\ \rho_\nu(\beta)  }{ S_{\beta=0}  },
		\end{equation}
		where $S_{\nu,d}$ is defined exactly as for the ground state, see Eq.~\eqref{eq:Sd_inf_time}, if we replace $|\psi\rangle$ by $|\nu\rangle$. Note that $S_\beta$ is {\em not} the normalized (eigenstate) diagonal entropy corresponding to the density matrix $\rho(lT)=\sum_\nu \rho_\nu |\nu(lT)\rangle \langle \nu(lT)|$. It is rather a measure of the average delocalization of the individual eigenstates of $H_F^{(0)}$ in the basis of the exact Floquet operator.
		\item (Eigenstate) energy density fluctuations $\overline{\delta \mathcal{E}}_\beta$ at finite-temperature:
		\begin{eqnarray}
		\overline{\delta \mathcal{E}}_\beta = \sum_\nu \delta \mathcal{E}_\nu\ \rho_\nu(\beta)  =  \sum_\nu \rho_\nu(\beta) \frac{1}{L}\sqrt{  \lim_{N_T\to\infty}\frac{1}{N_T}\sum_{l=0}^{N_T}\left(  \langle\nu(lT)| H_\text{F}^{(0)} |\nu(lT)\rangle   -  \langle\nu| H_\text{F}^{(0)} |\nu\rangle_d   \right)^2.   }
		\end{eqnarray}
		As with the entropy, $\overline{\delta \mathcal{E}}_\beta$ is not measuring density fluctuations in the system. Rather it measures the long-time fluctuations of the energy starting from a specific eigenstate and then averages over all eigenstates.
	\end{itemize}

	Let us now analyze the behavior of these observables in different driving regimes. Figure~\ref{fig:temperature} (a-c) shows the frequency dependence of the normalized energy $\overline{Q}_\beta$, the normalized (eigenstate) diagonal entropy $\mathcal{S}_\beta$ and the energy-density fluctuations $\overline{\delta\mathcal{E}}_\beta$ for various initial temperatures (see legend for details). Here, $2J$ sets the bandwidth of the lowest band of $H_F^{(0)}$, while $W_\mathrm{eff} = 2(J+J')$ - the total bandwidth of the two effective SSH bands. The bare hopping and bandwidth are denoted by $J_0$ and $W_0$, respectively. Fig.~\ref{fig:temperature} (a) shows the normalised energy of the system absorbed from the drive. Figure~\ref{fig:temperature} (b) illustrates the temperature dependence of the normalised (eigenstate) diagonal entropy. While at low frequencies all states heat up uniformly to infinite temperature, at large frequencies the states are only spread around the mean energy. Due to the high density of states in the middle of the spectrum, this spreading results in a higher (eigenstate) diagonal entropy than for the initial ground state. Finally, Fig.~\ref{fig:temperature} (c) shows the energy-density fluctuations as a function of temperature. Quite generally, it becomes visible that the size of the fluctuations decreases with increasing temperature. This effect is likely due to the additional statistical average involved. More interestingly, however, one sees that the high-frequency tail goes down significantly. Hence, the exponential decay of fluctuations as a function of the system size [see Fig.~\ref{fig:fluctuations} in the main text] is more pronounced for high-energy-density initial states in the high-frequency thermalising phase, which is expected from typicality.
	
	\begin{figure}[h!]
		\includegraphics[width=0.5\columnwidth]{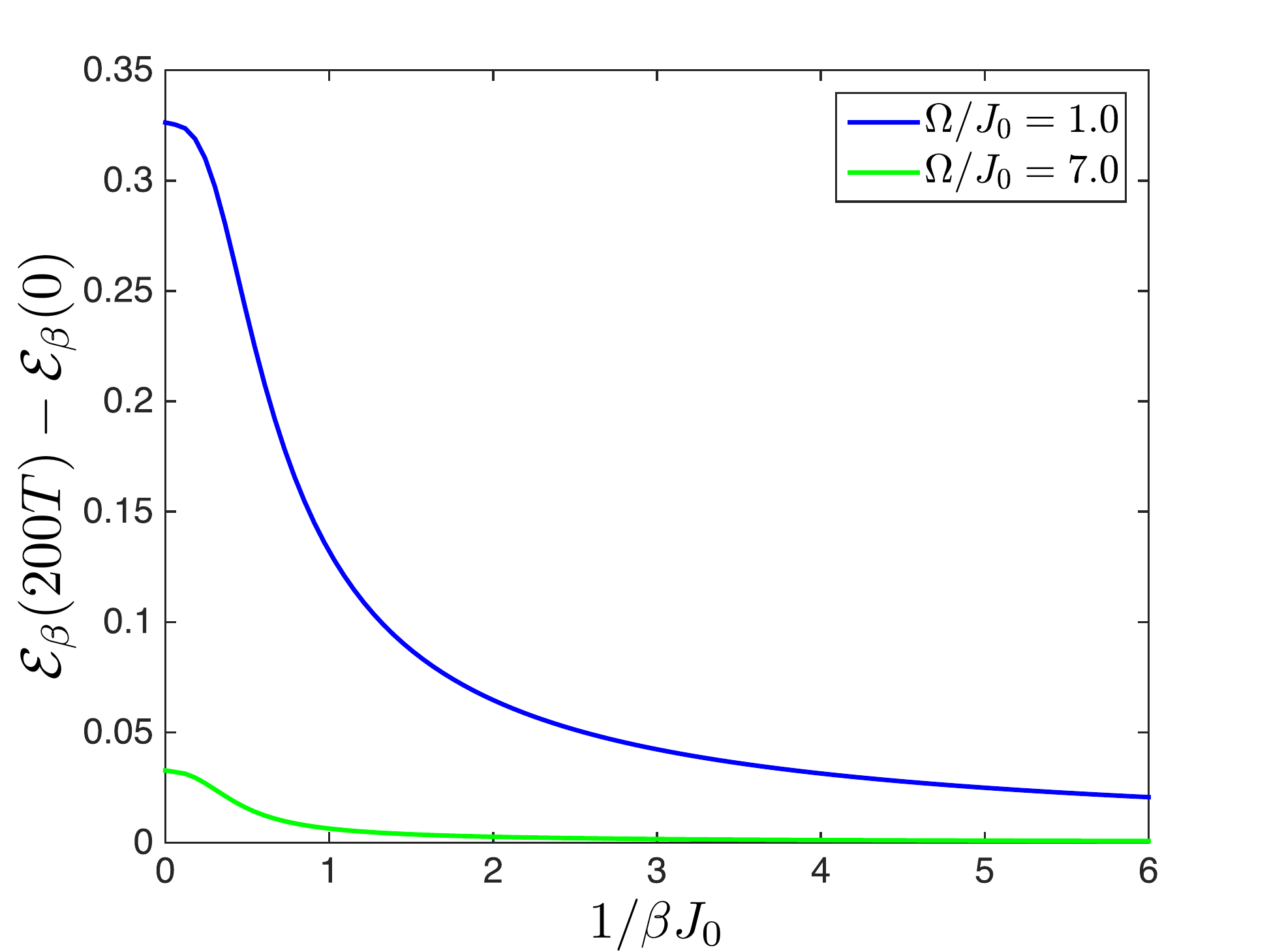}
		\caption{\label{fig:E_vs_temp} (Color online). Energy density $\mathcal{E}_\beta(200 T) - \mathcal{E}_\beta(0)$ pumped into the system as a function of the temperature of the Gibbs initial state localised around the GS. The parameters are $L=16$, $U/J_0=1$, $\zeta = 0.6$, $\delta\zeta = 0.12$, which amounts to $J'/J_0 = 0.41$, $J/J_0=0.29$.  }
	\end{figure}

	Last, in Fig.~\ref{fig:E_vs_temp} we also show the energy pumped into the system after the experimentally-relevant time scales of $200$ driving cycles of evolution, starting from a finite-temperature Gibbs state. We limit the discussion to high frequencies where the system does not heat up. For $\beta^{-1} = J$ the temperature is set within the lowest \emph{effective} band of $H_F^{(0)}$, but we can also consider other interesting cases where the temperature lies in the \emph{effective} band gap $\beta^{-1} = 2J$, or within the non-driven band $\beta = J_0$. Interestingly, one sees that higher-temperature initial states absorb less energy. Note also that, at low temperatures, the energy density absorbed from the drive decreases with increasing the drive frequency.

	\section{\label{app:Floquet corrections} Leading-Order Finite-Frequency Corrections}
	
	In this section, we calculate the leading $\Omega^{-1}$-corrections to the effective (drive-phase independent) Floquet Hamiltonian and the kick operator within van Vleck perturbation theory. We begin by casting the exact time-dependent rotating frame Hamiltonian in spin language via $S^-_m = a_m$ and $T^-_m = b_m$. The spin operators obey the spin-$1/2$ algebra $[S^-_m,S^+_n] = -2\delta_{mn}S^z_{m}$, $[T^-_m,T^+_n] = -2\delta_{mn}T^z_{m}$. Then the Hamiltonian in the rotating frame can be written as
	\begin{eqnarray}
	H^\text{rot}(t) &=& -J_0g(t)\sum_{m=1}^{L/2}\left( T^+_mS^-_m + \text{h.c.}\right)  -J_0h(t)\sum_{m=1}^{L/2-1}\left( S^+_{m+1}T^-_m + \text{h.c.}\right) +U  \sum_{m=1}^{L/2} S^z_mT^z_m +U  \sum_{m=1}^{L/2-1} S^z_{m+1}T^z_m,
	\end{eqnarray} 	
	where the functions $g(\tau)$ and $h(\tau)$ with $\tau = \Omega t$ are given by
	\begin{eqnarray}
	g(\tau) &=& e^{-i\left[\tau - (\zeta-\delta\zeta)F(\tau)\right]},\nonumber\\
	h(\tau) &=& e^{+i\left[\tau - (\zeta+\delta\zeta)F(\tau)\right]},\nonumber
	\end{eqnarray}
	\begin{equation}
	F(\tau) = \int f(\tau)\mathrm{d}\tau = \left\{ \begin{array}{ccc}
	\ \ \tau & \text{for} & -\pi/2\leq \tau\leq \pi/2 \\
	-\tau+\pi & \text{for} & \ \ \ \ \pi/2\leq \tau\leq 3\pi/2
	\end{array}\right.\nonumber
	\end{equation}
	
	Floquet's theorem applies to time-periodic Hamiltonians and reads
	\begin{eqnarray}
	U(t_2,t_1) = e^{-iK_\mathrm{eff}(t_2)}\; e^{-i(t_2-t_1) H_\mathrm{eff}}\; e^{iK_\mathrm{eff}(t_1)},
	\end{eqnarray}
	with the effective (non-stroboscopic) Hamiltonian $H_\mathrm{eff}$ and the time-periodic kick operator $K_\mathrm{eff}(t)$, whose $\Omega^{-1}$-corrections are calculated with the help of the van Vleck inverse-frequency expansion as~\cite{rahav_03,rahav_03_pra,goldman_14,bukov_14,goldman_14_res,eckardt_15,mikami_15}
	\begin{eqnarray}
	H_\text{eff}^{(1)} &=& \frac{1}{\Omega}\bigg\{ J_0^2 \sum_{m}c_{hh}\left(S^z_m - T^z_m  \right) + c_{gg}\left(T^z_m - S^z_{m+1}  \right) - J_0^2c_{gh}\sum_{m}\left(S^+_{m+1}T^z_mS^-_m - T^+_{m+1}S^z_{m+1}T^-_m + \text{h.c.}\right)  \bigg\},\nonumber\\
	K_\text{eff}^{(1)}(t=0) &=& \frac{1}{\Omega}\bigg\{  -J_0\sum_{m} \left( \kappa_- T^+_mS^-_m +  \kappa_+S^+_{m+1}T^-_m +\text{h.c.}\right)    \bigg\}.
	\end{eqnarray}
	The first-order correction contains a staggered potential term, and a correlated (interaction-dependent) hopping. The on-site staggered potential breaks the topological properties of the Floquet Hamiltonian, similarly to other one-dimensional Floquet topological insulators~\cite{iadecola_15_top}. Stroboscopic symmetry-protected topological phases have been studied extensively in Ref.~\onlinecite{iadecola_15_top}. If we set $\zeta_\pm = \zeta\pm\delta\zeta$, the affective coefficients governing the dynamics in the localised phase can be evaluated in a closed form for the periodic step drive:
	\begin{eqnarray}
	c_{gg}(\zeta_-) &=& \frac{1}{4\pi i}\int_0^{2\pi}\mathrm{d}\tau_1\int_0^{\tau_1}\mathrm{d}\tau_2\left[ \left(1 - \frac{\tau_1-\tau_2}{\pi}\right) \mathrm{mod}\; 2\pi\right]\bigg[g(\tau_1)[g(\tau_2)]^* - (\tau_1\leftrightarrow \tau_2) \bigg] \nonumber\\
	&=& \frac{1}{(\zeta_- - 1)} - 8\zeta_-^2\frac{\cos(\pi\zeta_-)+1}{\pi^2(\zeta_-^2-1)^3}, \nonumber\\
	c_{hh}(\zeta_+) &=& \frac{1}{4\pi i}\int_0^{2\pi}\mathrm{d}\tau_1\int_0^{\tau_1}\mathrm{d}\tau_2\left[\left(1 - \frac{\tau_1-\tau_2}{\pi}\right)\mathrm{mod}\; 2\pi\right]\bigg[h(\tau_1)[h(\tau_2)]^* - (\tau_1\leftrightarrow \tau_2) \bigg] \nonumber\\
	&=& - c_{gg}(\zeta_+),\nonumber\\
	c_{gh}(\zeta_-,\zeta_+) &=& \frac{1}{4\pi i}\int_0^{2\pi}\mathrm{d}\tau_1\int_0^{\tau_1}\mathrm{d}\tau_2\left[\left(1 - \frac{\tau_1-\tau_2}{\pi}\right)\mathrm{mod}\; 2\pi\right]\bigg[g(\tau_1)h(\tau_2) - (\tau_1\leftrightarrow \tau_2) \bigg] \nonumber\\
	&=& -4\frac{4\zeta_-\zeta_+(\zeta_+^2+\zeta_-^2-2)\cos\frac{\pi\zeta_-}{2}\cos\frac{\pi\zeta_+}{2}  - \pi(\zeta_-^2-1)(\zeta_+^2-1)(\zeta_-^2+\zeta_+^2 - \zeta_-\zeta_+ -1)\frac{\sin\frac{\pi(\zeta_--\zeta_+)}{2}}{\zeta_--\zeta_+}}{\pi^2(\zeta_-^2-1)^2(\zeta_+^2-1)^2},\nonumber\\
	\kappa_-(\zeta_-) &=& -\frac{1}{2}\int_0^{2\pi}\mathrm{d}\tau\left[\left(1-\frac{\tau}{\pi}\right)\text{mod} \ 2\pi \right] g(\tau)  \nonumber\\
	&=& -i \frac{4\zeta_-\cos\frac{\pi\zeta_-}{2}+\pi(\zeta_-^2-1)\left(1+\zeta_-\left(1-\sin\frac{\pi\zeta_-}{2}\right)\right)}{\pi(\zeta_-^2-1)^2},\nonumber\\
	\kappa_+(\zeta_+) &=& -\frac{1}{2}\int_0^{2\pi}\mathrm{d}\tau\left[\left(1-\frac{\tau}{\pi}\right)\text{mod} \ 2\pi \right] h(\tau) = -\kappa_-(\zeta_+).
	\end{eqnarray}
	The effective Hamiltonian and the effective kick operator are related to the stroboscopic Floquet Hamiltonian, which governs the dynamics at times integer multiples of the driving period, by $H_F[0] = e^{-iK_\mathrm{eff}(0)}H_\mathrm{eff}\; e^{iK_\mathrm{eff}(0)}$, where the square bracket $[\cdot]$ denotes the Floquet gauge (or equivalently the initial phase of the drive), see Ref.~\onlinecite{bukov_14}.
	
	\begin{figure*}[h!]
		\includegraphics[width=\columnwidth]{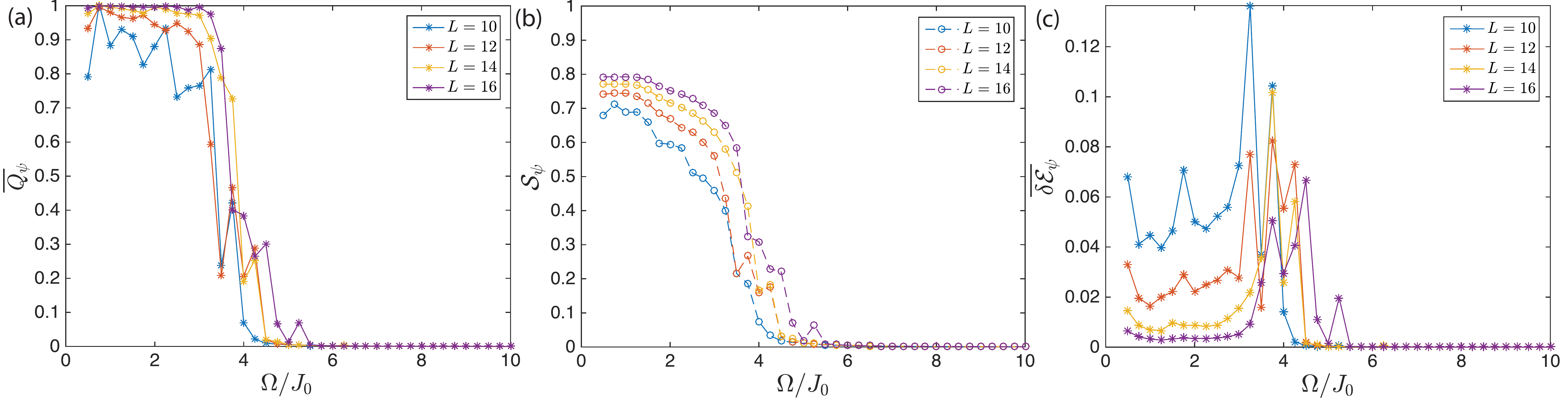}
		\caption{\label{fig:inf_times_corrected} (Color online). Frequency dependence of the normalized energy pumped into the system at infinite times $\overline{Q}_\psi$ (a), the diagonal entropy $\mathcal{S}_\psi$ (b), and the energy-density fluctuations $\overline{\delta\mathcal{E}}_\psi$ (b), starting from the ground state of the corrected Floquet Hamiltonian $H_\mathrm{eff}^{(0)}+H_\mathrm{eff}^{(1)}$, properly brought back to the lab frame by the leading-order kick operator $K_\mathrm{eff}^{(1)}(0)$. The parameters are $U/J_0=1$, $\zeta = 0.6$, $\delta\zeta = 0.12$, which leads to $J/J_0 = 0.41$ and $J'/J_0=0.29$. }
	\end{figure*}
	
	When included, the leading correction term is expected to reduce the energy injected into the system in the high-frequency tail by suddenly starting the drive. To test this, we start from the ground state of the Hamiltonian $H_\mathrm{eff}^{(0)}+H_\mathrm{eff}^{(1)}$, appropriately rotated back to the lab frame by the kick operator $K_\mathrm{eff}^{(1)}(0)$, and simulate the normalised energy at infinite times, and the diagonal entropy as shown the result in Fig.~\ref{fig:inf_times_corrected}. When compared to the curves in Fig.~\ref{fig:U_dep} of the main text, we see that, while the small-frequency behaviour leading to heating to infinite temperature remains qualitatively the same, the energy injected into the system due to suddenly starting the drive at time $t_0=0$ becomes negligible, as expected. This check is important, as experiments are always performed at finite frequencies.

	\section{\label{app:level_statistics} Level Statistics}
	
	One of the standard measures of ergodicity in quantum systems is the level spacing statistics.  According to Random Matrix Theory, ergodic Hamiltonians are well-described by the Gaussian Orthogonal Ensemble (GOE) with their level spacing statistics following the Wigner-Dyson distribution. For non-ergodic Hamiltonians, on the other hand, one expects a Poisson distribution. In general, it is believed that there exists a one-to-one correspondence between Wigner-Dyson distributed level spacings of a quantum model and chaotic dynamics in the classical limit~\cite{bohigas_84}. Periodically-driven systems feature the additional subtlety that quasienergies are defined only modulo multiples of the driving frequency. In this respect, it has been shown that the level statistics of the approximate Hamiltonian obtained via the inverse-frequency expansion is not a good measure of ergodicity, since the folding of the many-body spectrum can introduce artificial correlations in the level spacings. This is intimately related to the fact that the inverse-frequency expansions do not capture any photon-absorption resonances~\cite{pweinberg_15}, and the hybridisation of the corresponding levels. Nevertheless, the folded spectrum of the exact Floquet Hamiltonian can still be used to extract useful information about ergodicity of the underlying dynamics~\cite{dalessio_14}. The classification of the symmetry classes allowed for the Floquet Hamiltonian in the presence of disorder has been studied in Ref.~\cite{regnault_15}.
	
	\begin{figure*}[h!]
		\includegraphics[width=\columnwidth]{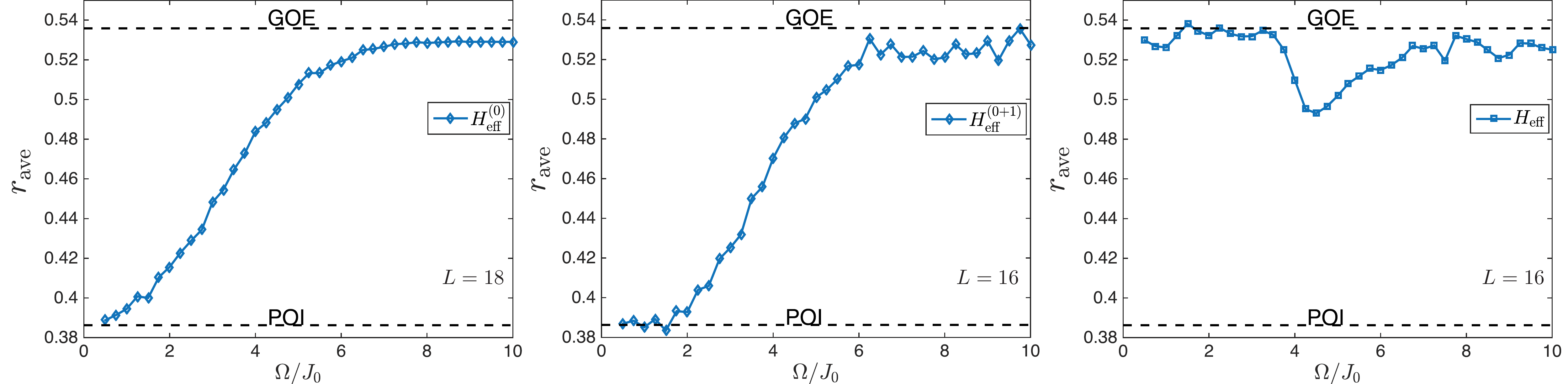}
		\caption{\label{fig:level_statistics} (Color online). Frequency-dependence of the mean level spacing $r_\mathrm{ave} = \mathrm{min}(\delta_{i+1}, \delta_i)/\mathrm{max}(\delta_{i+1},\delta_i)$ in the spectra of the infinite-frequency Hamiltonian $H_\mathrm{eff}^{(0)}$ describing the interacting SSH model (a), the corrected Floquet Hamiltonian to leading order $H_\mathrm{eff}^{(0+1)} = H_\mathrm{eff}^{(0)}+H_\mathrm{eff}^{(1)}$ (b), and the exact Floquet Hamiltonian $H_\mathrm{eff}$ (c). The dashed horizontal $U/J_0=1$, $\zeta = 0.6$, $\delta\zeta = 0.12$, which amounts to $J'/J_0 = 0.41$, $J/J_0=0.29$.}
	\end{figure*}
	
	Studying the level statistics of a Hamiltonian requires a careful binning of the data. Fortunately, the mean level spacing $r_\mathrm{ave} = \mathrm{min}(\delta_{i+1}, \delta_i)/\mathrm{max}(\delta_{i+1},\delta_i)$ where the phases $\delta_i = (E_F^{i+1} - E_F^i)T$ already contain the necessary information to reveal the statistics of the level spacings: if $r_\mathrm{ave} = 0.5358$, the level statistics is Wigner-Dyson, whereas if $r_\mathrm{ave} = 0.3862$ -- it is Poisson distributed. Figure~\ref{fig:level_statistics} shows $r_\mathrm{ave}$ as a function of frequency for the infinite-frequency Floquet Hamiltonian $H_\mathrm{eff}^{(0)}$ (a), the leading correction $H_\mathrm{eff}^{(0)}+H_\mathrm{eff}^{(1)}$ (b), and the exact Floquet Hamiltonian $H_\mathrm{eff}$ (c). We would like to make a few remarks: (i) it becomes clear that ergodicity at infinite-frequencies is indeed fully attained, due to the drive-engineered small level of dimerisation of the chain, which renders the model non-integrable. This is correlated with the presence of Wigner-Dyson statistics of the spectrum at high-frequencies. Including the leading-order finite-frequency correction, which features interaction-dependent hopping terms, does not change the level spacing. (ii) at intermediate-to-low frequencies, the level statistics of the inverse-frequency expansion is messed up due to the folding of the spectrum which influences the level spacings in an artificial way. Our results are in full agreement with those in Ref.~\onlinecite{dalessio_14}. (iii) the level statistics of the exact Floquet Hamiltonian features Wigner-Dyson statistics both at high and low frequencies [as expected for a system featuring thermalising dynamics], while a clear dip is visible in the crossover regime, signalling non-thermal statistics. This is yet another evidence for the glassy dynamics observed at intermediate frequencies.

\end{widetext}
\end{appendix}

\end{document}